\documentclass[ag]{copernicus}

\usepackage{color}

\begin{document}

\title{{Electron-cylotron maser radiation from electron holes: \\ Upward current region}}

\author[1,2]{{R. A. Treumann}
}
\author[3]{{W. Baumjohann}}
\author[4]{{R. Pottelette}}

\affil[1]{Department of Geophysics and Environmental Sciences, Munich University, Munich, Germany}
\affil[2]{Department of Physics and Astronomy, Dartmouth College, Hanover NH 03755, USA}
\affil[3]{Space Research Institute, Austrian Academy of Sciences, Graz, Austria}
\affil[4]{LPP-CNRS/INSU, 94107 Saint-Maur des Foss\'es, France}

\runningtitle{Electron-Hole Radiation}

\runningauthor{R. A. Treumann, W. Baumjohann, and R. Pottelette}

\correspondence{R. A.Treumann\\ (rudolf.treumann@geophysik.uni-muenchen.de)}

\received{ }
\revised{ }
\accepted{ }
\published{ }


\firstpage{1}

\maketitle

\begin{abstract}
Electron holes are suggested to be an important source for generation of electron-cyclotron maser radiation. We demonstrate that electron holes generated in a ring-horseshoe distribution in the auroral-kilometric radiation source region have the capacity to emit  band-limited radiation. The radiation is calculated in the proper frame of a circular model hole and shown to be strictly perpendicular in this frame. Its bandwidth under auroral conditions is of the order of $\sim1$ kHz, which is a reasonable value. {It is also shown that much of the drift of fine structure in the radiation can be interpreted as Doppler shift. Estimates based on data are in good agreement with theory. Growth and absorption rates have been obtained for the emitted radiation. However, the growth rate of a single hole obtained under conservative conditions is small, too small for reproducing the observed fine structure flux. Trapping of radiation inside the hole for the hole's lifetime helps amplifying the radiation additionally but introduces other problems. This entire set of questions is discussed at length and compared to radiation from the global horseshoe distribution.} The interior of the hole produces a weak absorption at slightly higher frequency than emission. The absorptivity is roughly two orders of magnitude below the growth rate of the radiation thus being weak even when the emission and absorption bands overlap. Transforming to the stationary observer's frame it is found that the radiation becomes oblique against the magnetic field. For approaching holes the radiated frequencies may even exceed the local electron cyclotron frequency.

\end{abstract}

\introduction
It is common view that the electron-cyclotron maser acts on the global anisotropic (or loss-cone) electron phase-space distribution \citep[more precisely the relativistic momentum distribution, cf., e.g.,][for a review see Treumann, 2006]{wulee1979} and that structures which are localised in momentum space, so-called fine-structures, have no or little effect on the generation of radiation \citep[for a review cf., e.g.,][and references therein]{louarn2006}. 

\begin{figure}[t!]
\centerline{{\includegraphics[width=0.5\textwidth,clip=]{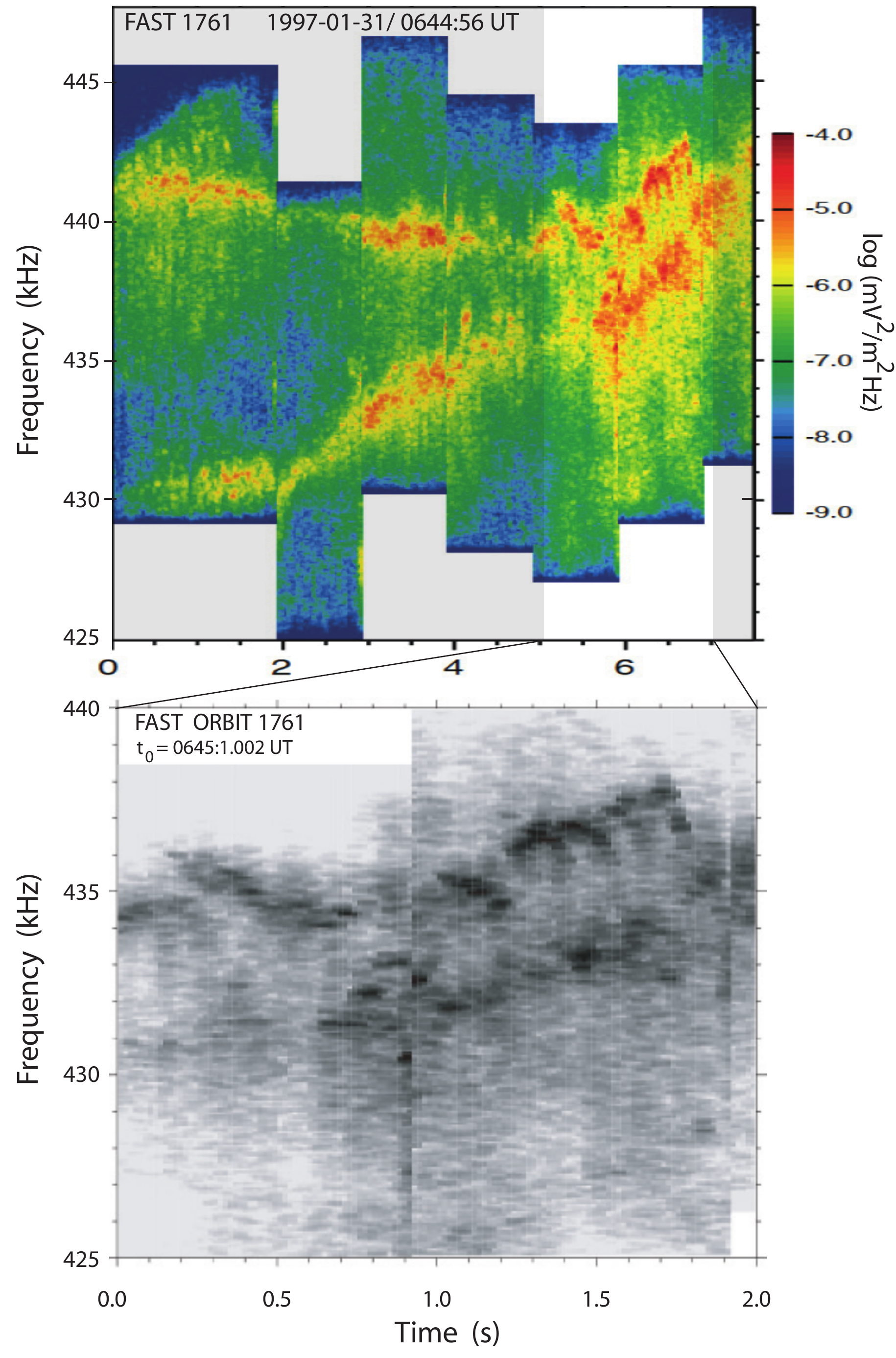}
}}
\caption[ ]
{\footnotesize {An example of two drifting auroral kilometric radiation emission bands ({\small FAST} orbit {\small 1761}) consisting of narrow fine structures. The expanded high-time resolution dynamic spectrum of the unshaded part of the spectrum is shown in the lower panel \citep[re-scaled data taken from][]{pottelette2000}. Time resolution is $\Delta t= 33$ ms. The two approaching emission bands consist of many individual `elementary radiation structures'. The upper band shows towards the end of the observation shows indications of turnovers of some of the elementary structures. The limited time resolution does not allow for an unambiguous determination of the individual frequency drifts. In the turnovers the frequency drift is zero. Simultaneously, the spectral energy density maximises. The highest   power flux in those single 1 kHz bandwidth events amounts to ${\cal P} \sim 10^{-7}$ W/m$^2$ (or energy density $W\sim10^{-15}$ J/m$^3$). Since the emission frequency should be related to the location of the radiation source along the magnetic field via the electron cyclotron frequency, the drift of the  two bands suggests reflection at a layer located at higher magnetic field strengths and is moving slowly downward.}}\label{fig-akr-spectrum}
\vspace{-0.3cm}
\end{figure}

\begin{figure}[t!]
\centerline{{\includegraphics[width=0.5\textwidth,clip=]{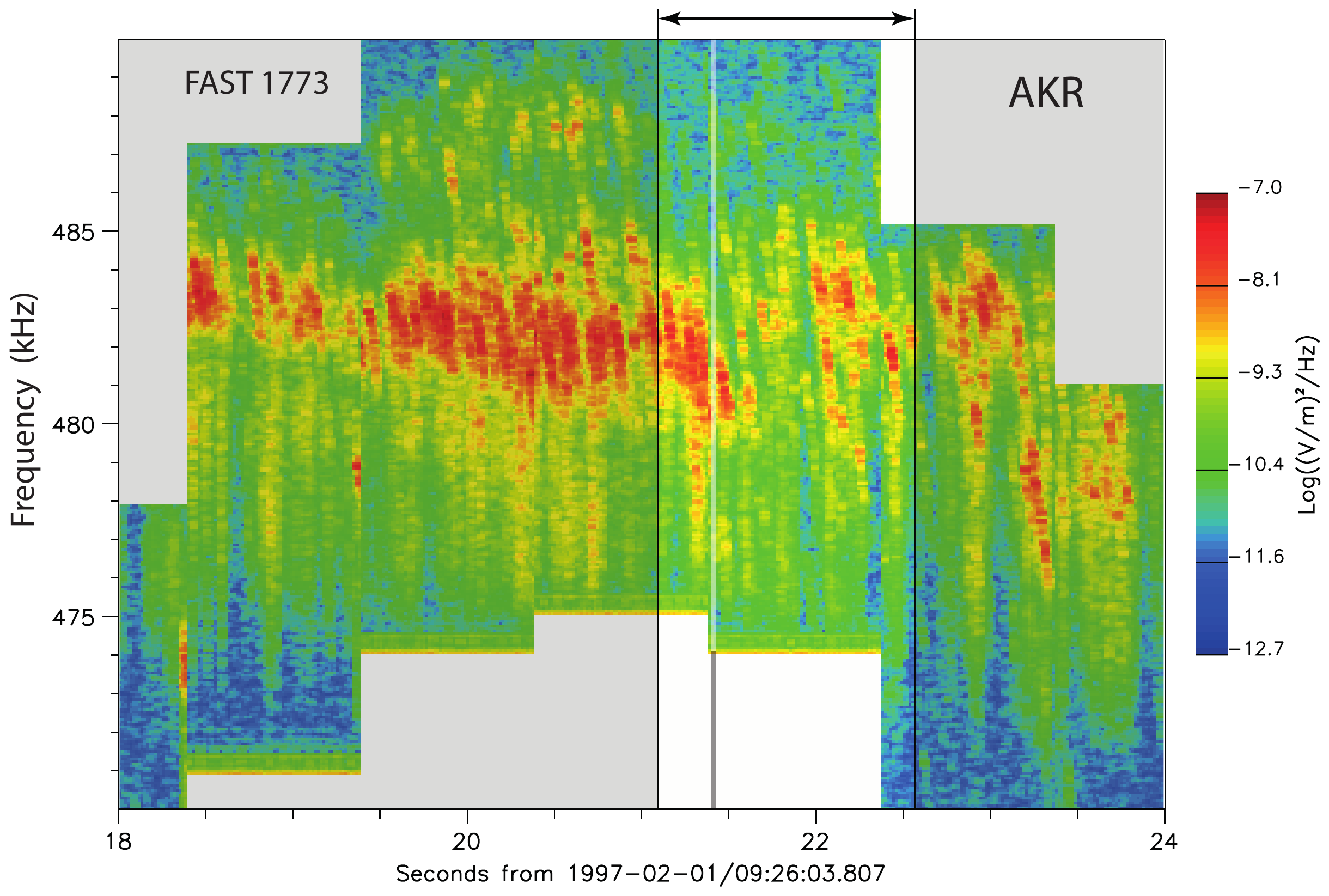}
}}
\caption[ ]
{\footnotesize {Six second snapshot of power spectral density of Auroral Kilometric Radiation in the highest available resolution ({\small FAST TRACKER} instrument). The slice between the two vertical bars indicated by the two-sided arrow corresponds to the shaded slice in Figure \ref{fig-over} and the fully resolved electron data of Figure \ref{fig-electrons-eh}. The highest available time resolution is $\sim$30 ms. The radiation consists of many short lived radiation events of $\sim$(1-2)\% bandwidth. The highest power flux in these events amounts to ${\cal P}\sim 10^{-11}$ W/m$^2$ (or energy density $W\sim10^{-19}$ J/m$^3$). In this example the events exhibit negative frequency drifts of $df/dt\sim-100$ kHz/s with the implication that in this particular case the individual radiation sources move into a region of decreasing magnetic field strength, i.e. upward along the geomagnetic field at velocity $10^3\lesssim V\lesssim 10^4$ km/s. The shaded narrow vertical slice corresponds to the low-frequency waveform snapshot in Figure \ref{fig-electrons-eh}.}}\label{fig-rad}
\vspace{-0.3cm}
\end{figure}

What, however, if electron holes cause a distinct structure not only in real (configuration) space but also in the electron-momentum space distribution? Could they contribute to the generation of radiation? Quite generally, electron holes have indeed been suggested to be an important radiation source as well via the electron-cyclotron maser mechanism. This suggestion was based on  observations of auroral kilometric-radiation spectra with high time and frequency resolution \citep{pottelette2000,pottelette2007} {and on the analysis of low-frequency electrostatic waves in the auroral upward current region, i.e. the auroral kilometric radiation source \citep{pottelette2005}. Such spectra are reproduced in Figures \ref{fig-akr-spectrum} and \ref{fig-rad} for two different {\small FAST} spacecraft passages across the auroral kilometric radiation source region. Data obtained during the second of these passages will serve as a guide for the present investigation.}  

These observations even led to the further suggestion  \citep{pottelette2000} that, possibly, all of the auroral kilometric radiation could be made up by electron-cyclotron maser radiation emitted from electron holes drifting at high speed through space, with the frequency drift mapping the variation of the electron-cyclotron frequency  along the magnetic flux tubes to which they are confined. {Though such a very strong and exclusive interpretation might occasionally be realised during weak radiation events, we do neither insist on nor refer to it. The present communication just intends to offer a simple mechanism of how \emph{single} electron holes could be involved into generation of electron cyclotron maser radiation.}

{The pre-requisite for the electron-cyclotron maser to become active is the existence of perpendicular phase space gradients, i.e. positive derivatives of the electron distribution function $f_e(v_\|,v_\perp)$ with respect to the perpendicular velocity $v_\perp$}  \citep{pottelette2000,pottelette2005}. Recently, this condition has been taken for investigating the bending of an electron hole in phase space \citep{treumann2008} under the assumption that the electron hole is a narrow entity both in configuration and phase space. {Here we calculate the electron cyclotron maser emission from an electron hole under the assumption of a finite extension of the hole in $v_\perp$, independent of a possible bending of the hole.  Just a brief argument will be given for justifying the model of the electron hole used in the calculation.}

\section{{A Brief Review of Relevant Observations}}
\begin{figure}[t!]
\centerline{{\includegraphics[width=0.5\textwidth,height=0.7\textheight,clip=]{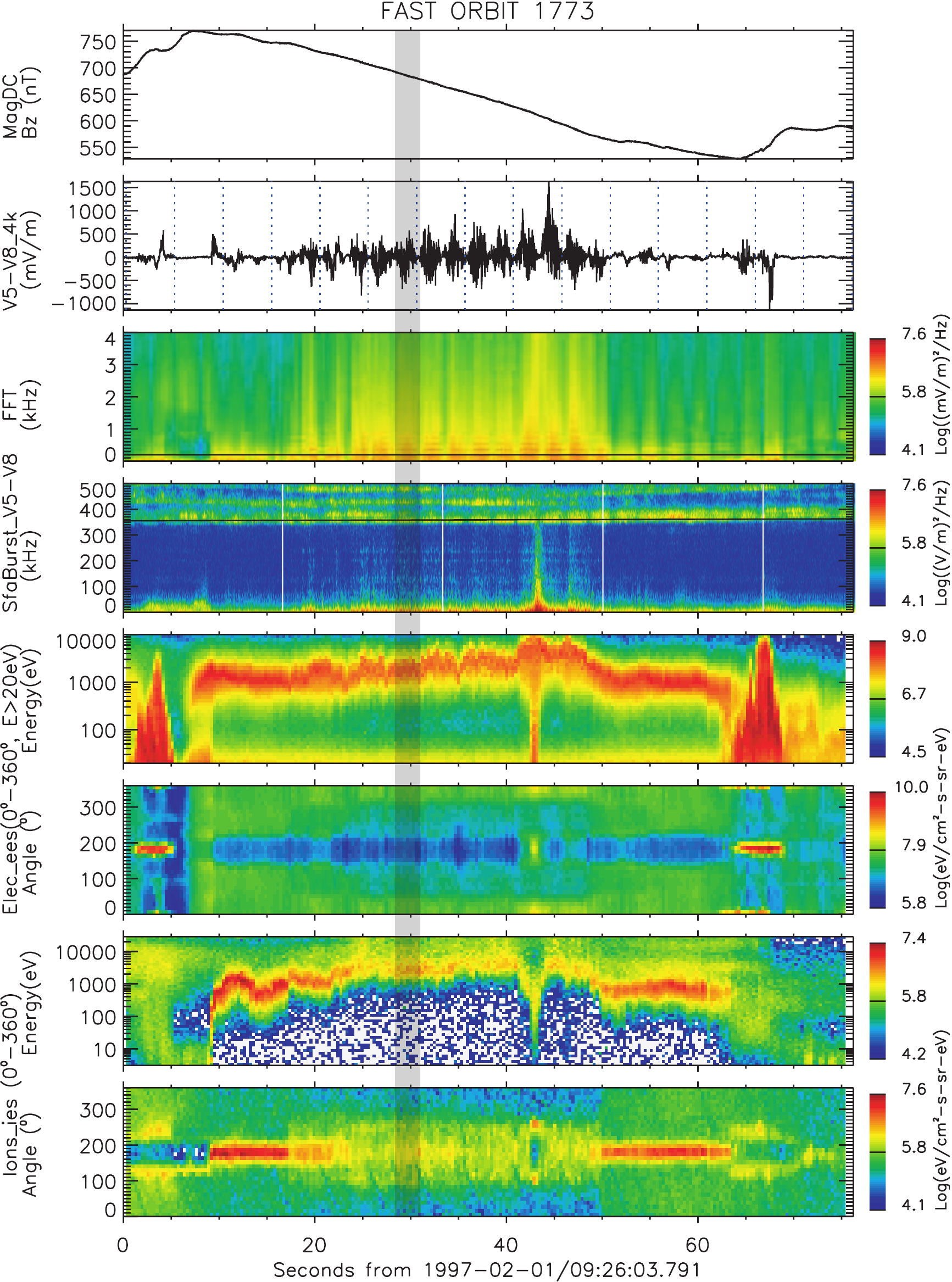}
}}
\caption[ ]
{\footnotesize {A {\small FAST} spacecraft passage across the auroral kilometric radiation source region showing from top to bottom the magnetic component $B_z$ with its typical sheet current slope; the highly variable low-frequency electric wave amplitude reaching amplitudes of $|\delta E|\gtrsim \pm0.5$ V/m; low frequency $f<4$ kHz broadband electric field emissions typical for electron holes when exceeding the plasma frequency $f_{e}=\omega_e/2\pi\lesssim 2$ kHz; three cyclotron harmonic emissions in the AKR spectrum above the electron cyclotron frequency (black horizontal); the spectral electron energy density; angular distribution of electron fluxes; ion spectral energy density; and the ion angular distribution. The shaded slice is shown in highest resolution in the following figures (data obtained within the French-UCBerkeley co-operation).}}\label{fig-over}
\vspace{-0.3cm}
\end{figure}
{Before turning to radiation theory we add a comprehensive account of  relevant simultaneous field and particle observations.} 

{Figure \ref{fig-over} gives an overview of a {\small FAST} spacecraft passage through the auroral kilometric radiation source region. Such data have been shown earlier \citep[for instance][and others]{ergun1998a,ergun1998b,pottelette2000}. The data are exhaustively described in the caption.} 

{The top panel shows the magnetic field of the typical sheet current profile across the extended upward current carried by the downward several-keV electrons (fifth panel) with angular horseshoe distribution (sixth panel). The horseshoe  is characterised by lacking electrons at pitch angles $180^\circ\pm20^\circ$ in the loss cone and maximum electron fluxes around 0$^\circ$ and 360$^\circ$. The energy spread of the auroral electron distribution, i.e. the horseshoe electron temperature, is not much less than the average electron energy.} 

{The second to fourth panels show spectral wave observations: highly fluctuating total electric wave amplitudes in panel two reaching up to 1 V/m, which are related to the broadband low-frequency emissions in panel three and being typical for the signatures of a large number of electron holes along the path of the spacecraft; these emissions are overlaid by equally spaced ion-cyclotron harmonics which coincide with ion conics in the two lowest panels showing the ion flux. These are of little interest here. }

The fourth panel shows the overview of the auroral kilometric radiation at and above the electron cyclotron frequency {$\omega_{ce}/2\pi$}, the black horizontal line at $\sim$360 kHz. Three electron cyclotron-harmonic bands are covered. One notices the well-known extension of the radiation below the electron cyclotron frequency in the entire radiation source region.
 
{Here we just summarise some relevant facts in panels 2, 3, 5 and 6 from top:} 

\noindent{--- Panel 5 in the kilometric radiation source (roughly between 10 and 60 s) shows the electron energy flux forming an intense band in the keV range with energy spread less but comparable to central electron energy.}

\noindent{--- Panel 6 shows the angular distribution of these fluxes to concentrate around 0$^\circ$ and 360$^\circ$ but spreading in angle with broad completely empty loss cone around 160$^\circ$ to 200$^\circ$. Note the steep decrease of the fluxes with increasing angle. Such spectra are typical for horseshoe-ring distributions.}

\noindent{--- Panel 2 shows electric wave forms reaching large amplitudes up to low-frequency electric wave fields of $0.5$ to $1.0$ V/m being strongly modulated not only by spacecraft spin (the 4 s modulation) but predominantly intrinsically.} 

\noindent{--- Panel 3 shows the corresponding low frequency wave spectrogram exhibiting very broad band temporarily highly variable signals with frequency far exceeding the plasma frequency.  Electron densities (not shown) range here, in the `auroral cavity', around 0.05 cm$^{-3}$ yielding plasma frequencies $\omega_e/2\pi\lesssim$ 2 kHz. The horizontal black line at few 100 Hz is the ion-cyclotron frequency.}

\noindent{--- Such wave spectra correspond to the spacecraft crossing many closely spaced narrow electric wave structures with steep boundaries. They belong to electron holes \citep[cf.][for the proof of the existence of electron holes in the auroral kilometric radiation source]{pottelette2005}.} 

\noindent{--- One may note that these spectra also show signs of  equally spaced ion cyclotron harmonics which are the result of the known instability of electron holes \citep[cf., e.g.,][]{oppenheim2001} with respect to these waves.} 

{A six second snapshot of free space high frequency ($\omega\gg\omega_e$) electromagnetic wave spectra taken from the same spacecraft orbit and time-slot accidentally caught by the {\small FAST TRACKER} instrumentation is reproduced in the highest available temporal and frequency resolution in Figure \ref{fig-rad} (one should note that resolution is limited by the `Fourier uncertainty relation' $\Delta\omega\Delta t\sim 1$). Temporal resolution is $\Delta t\sim 33$ ms.} 

{This figure shows an intense narrow (total bandwidth $\lesssim 10$ kHz corresponding to $\Delta\omega/\omega\lesssim 2$\%) auroral kilometric radiation band emitted in the X-mode and being structured by a sequence of many short radiation pulses of similar though not identical negative spectral drift $\Delta \omega/\Delta t<0$ and with highest intensities being restricted to a bandwidth of $\sim$1 to 2 kHz only. Drifts range around 100 kHz/s.} 

\begin{figure}[t!]
\centerline{{\includegraphics[width=0.5\textwidth,clip=]{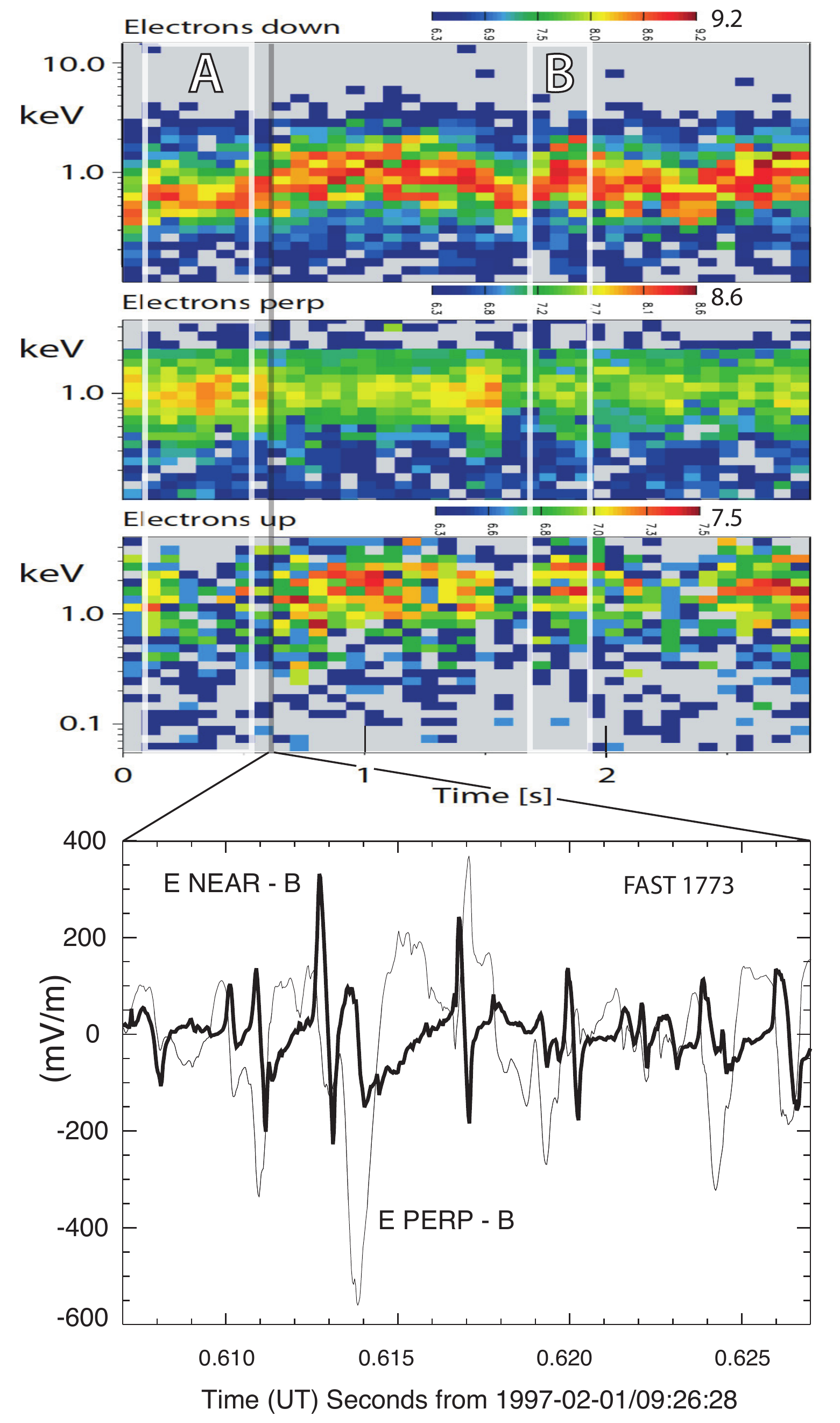}
}}
\caption[ ]
{\footnotesize{Electron energy fluxes and waveform data from the shaded slice in Figure \ref{fig-over} shown in the highest available resolution. The two stripes labelled A and B have been used for determining average distributions in Figure \ref{fig-dist}. \emph{Top panel}: Downward electrons  (within an angle of $\pm11^\circ$ parallel to the magnetic field) The energy flux in the most intense events is $\sim 6.4\times10^{-4} \Delta\epsilon$ J/m$^2$s, with $\Delta\epsilon\sim 1-3$ (in keV) the bandwidth of the electron energy. \emph{Second panel}: perpendicular electrons. \emph{Third panel}: Upward (reflected) electrons. Note the differences in the colour bars. Upward electron fluxes are roughly two orders of magnitude weaker than downward. Perpendicular (trapped) electron fluxes are about one order of magnitude weaker taking into account that the colours do not reach saturation. \emph{Bottom panel}: A 20 ms snapshot of low-frequency electric wave form parallel and perpendicular to the magnetic field showing the bipolar parallel and unipolar perpendicular electric fields of several electron holes crossing the {\small FAST} spacecraft. Their temporal width is $\Delta t \sim 5$ ms. These structures cannot be resolved by the electron instrument whose time resolution is roughly $\sim$100 ms.}}\label{fig-electrons-eh}
\end{figure}

{Some internal structure may be indicated down to the 33 ms time resolution, but the intense emissions are close to saturation, and the reality of the finest structures of $\sim$ 0.1 to 0.2 kHz in frequency is questionable; their proximity to the ion cyclotron frequency is intriguing and might point to some kind of modulation of the radiation though. This is of less interest in the context of this paper but might be related to observations of modulation of solar radio called `Zebra bursts' \citep{treumann2011a}. The non-shaded window in this figure corresponds to a snapshot of highest resolution electron flux data shown in Figure \ref{fig-electrons-eh}.}

\begin{figure}[t!]
\centerline{{\includegraphics[width=0.45\textwidth,clip=]{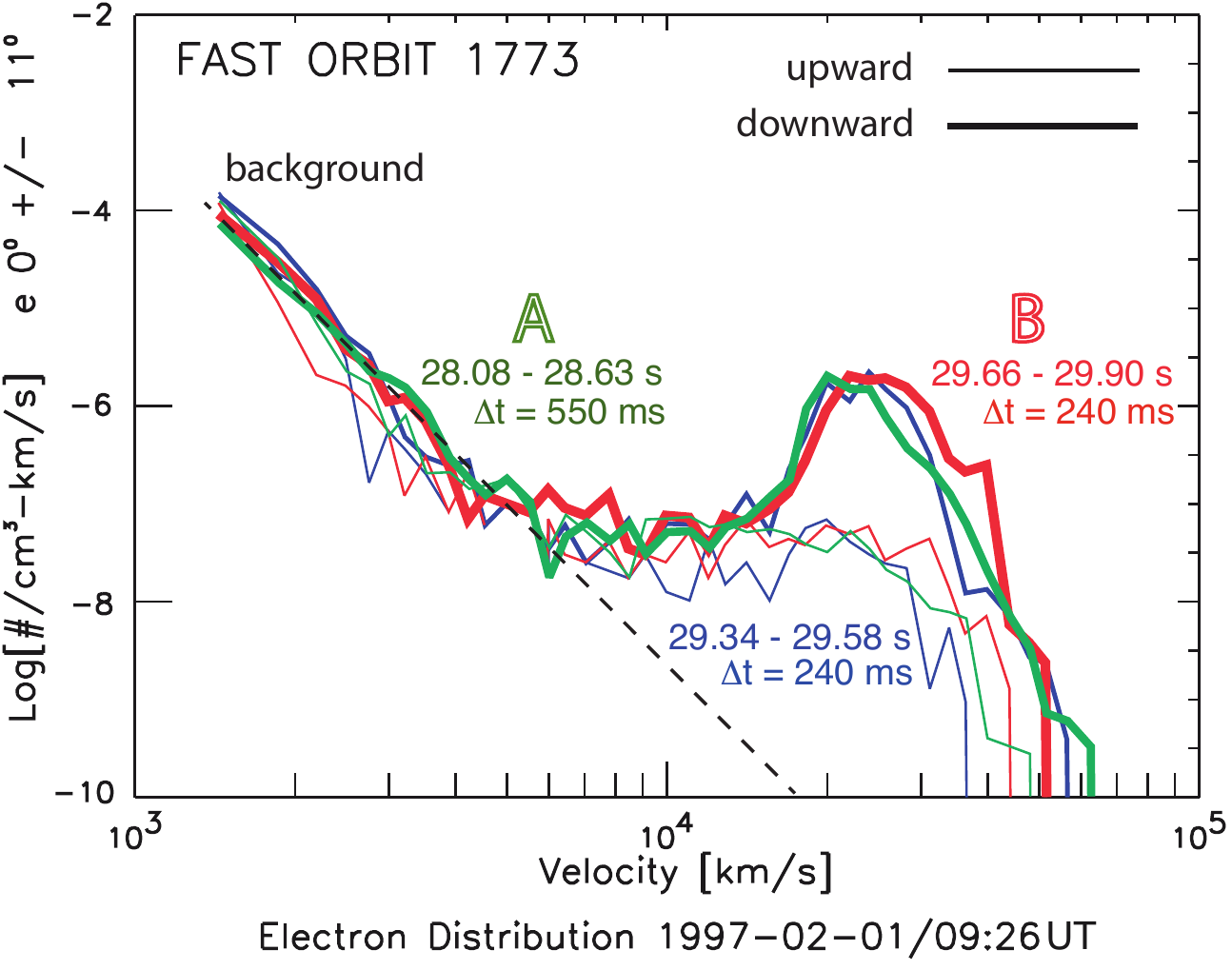}
}}
\caption[ ]
{\footnotesize{The magnetic-field aligned electron velocity distribution in the auroral-kilometric radiation source above active aurorae during the observation time of Figures \ref{fig-over} - \ref{fig-electrons-eh} in the downward electron flux (upward current) region. Observation times are given at the curves. The dashed line indicates background fluxes.  Upward (reflected) electrons (thin lines) and downward electrons (heavy solid lines) belong to three different time slices in Figure \ref{fig-electrons-eh}. The two interesting slices are labelled A and B. Integration times $\Delta t$ are indicated. These observations are typical for a downward  ring-horseshoe distribution with highest flux down along the magnetic field (panel 6 in Fig. \ref{fig-over}). Clearly, the limited resolution does not resolve any electron holes. The electron distributions are cut off at high velocity. The highest velocity peak (caused by the two saturated high-energy red spots in slice B of Figure \ref{fig-electrons-eh}) at the highest energy parallel ring distribution B (red curve) may be due to formation of the accelerated cold electron beam which forms in the evolution of the electron hole (at plasma frequency of $\omega_e/2\pi\approx (1.5-2)\times 10^3$ Hz the integration time of 0.24 s corresponds to $\approx 500$ plasma periods, sufficient time for formation of holes and generation of the cold fast beam). This beam has maximum velocity $V_b\sim 4.5\times 10^4$ km\,s$^{-1}\sim 0.15c$. The velocity of the ring-horseshoe is $V_R\sim 2.5\times10^4$ km\,s$^{-1}$. Its `thermal' velocity spread is about $\Delta V=v_e\sim 10^4$ km\,s$^{-1}$.}}\label{fig-dist}
\vspace{-0.3cm}
\end{figure}

{Three seconds of electron data measured by good fortune in highest available resolution during the high-resolution radiation observations of Figure \ref{fig-rad} shown in Figure \ref{fig-electrons-eh} are divided into three groups: downward auroral current electrons at pitch angles $0^\circ\pm11^\circ$ in the top panel, upward mirrored (magnetically trapped) electrons with pitch angles $180^\circ\pm20^\circ$ in the bottom panel, and perpendicular electrons with pitch angles $90^\circ\pm15^\circ$ in the middle panel. Mirrored upward electrons have pitch angles outside the loss cone being of slightly higher energy but much lower fluxes. All electron fluxes are modulated in energy and intensity. The two slices labelled A and B are later used for determination of reduced downward and upward distribution functions in Figure \ref{fig-dist}.}

{Auroral downward electrons exhibit quite intense fluxes, in a few cases even reaching into saturation. On the other hand, perpendicular electron fluxes never reach saturation even though their colour bar is set more than half a magnitude lower than for the downward electrons. Thus, perpendicular electron fluxes are at least one order of magnitude lower than downward fluxes. This observation, though not of importance for electron holes, is crucial in the calculation of the cyclotron maser radiation from a horseshoe-ring distribution.}

{The lower part of Figure \ref{fig-electrons-eh} shows a 20 ms snapshot of electric field wave forms (panel two in Figure \ref{fig-over}) parallel labelled E near-B) and perpendicular (labelled E perp-B)  to the magnetic field. The time slice to these measurements is the dark vertical line across the upper three panels in this figure. The behaviour of these wave forms is typical for the entire downward electron (upward current) time period in Figure \ref{fig-over}. The reason for choosing the above slice was that only during this particular slice in the entire sequence of high resolution observations the antennas were  directed approximately parallel and perpendicular with respect to the magnetic field direction, thus yielding the clearest picture of the spatial electric field structure.}

{Inspection of this short though representative sequence of wave forms shows the presence of a chain of short scale ($\sim$ few ms) bi-polar or non-symmetric quadru-polar electric field structures passing the spacecraft along the magnetic field and obeying a uni-polar or at most bipolar perpendicular electric field structure and reaching parallel amplitudes in this slice of $\sim$300 mV/m. These electric signals are typical for electron holes and, when being Fourier transformed, account for the extremely broad band signals in the low-frequency electric wave spectra in the third panel of Figure \ref{fig-over}. Clearly, with its $\sim$100 ms time resolution the electron instrument has no chance to resolve any of those structures. If one would be very brave one could interpret some yellow or even green spots in the red band in the top panel as signatures of an accumulation of electron holes on the electron energy flux which, however, would be highly speculative, and we refrain from it. At the currently available resolution of electron instrumentation on spacecraft the resolution of electron holes in the auroral kilometric radiation source in the electron distribution is illusory. }

{Reduced parallel distribution functions obtained from the electron measurements in the two time slices of Figure \ref{fig-electrons-eh} are plotted in Figure \ref{fig-dist}. There the heavy lines belong to downward fluxes while the light lines show reflected magnetically trapped electron distributions. The full distribution has been used such that at low velocities `photo'-electrons (labelled `background') around the spacecraft pollute the observations. The combination of upward and downward reduced distributions is typical for a ring-horseshoe electron distribution caused by the well-known combination of the mirror force and accelerating local parallel electric field on the electrons in the auroral kilometric radiation source region. These distributions have been obtained by integration over the indicated times (temporal slices).}
\begin{figure}[t!]
\centerline{{\includegraphics[width=0.45\textwidth,clip=]{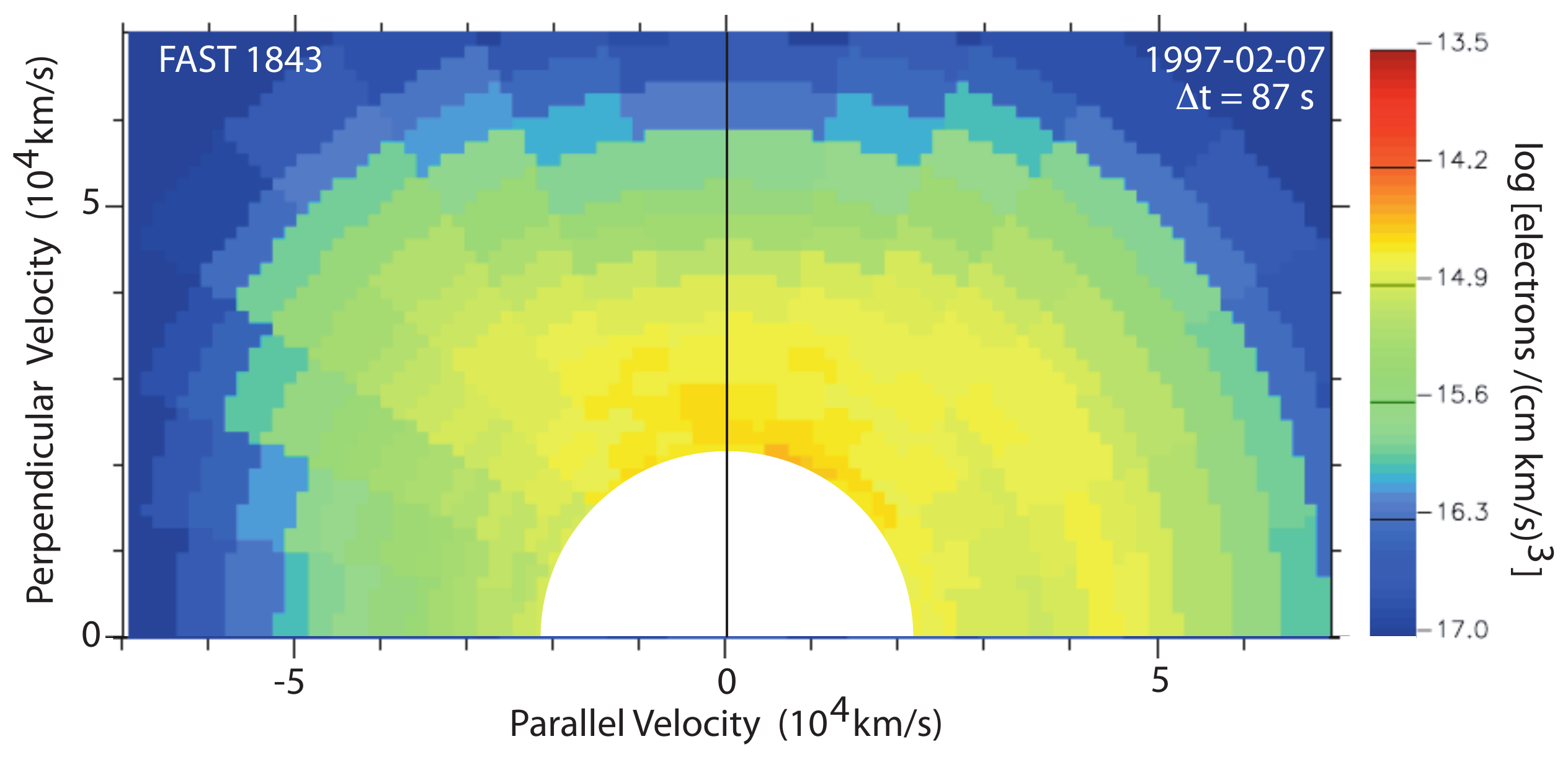}
}}
\caption[ ]
{\footnotesize{A typical full ring-horseshoe distribution of electrons measured within a time interval of $\Delta t=87$ s during {\small FAST 1843}. The low-energy electron background has been eliminated. The distribution is plotted in $v_\perp,v_\|$ space. The relative intensities are given by the colour bar on the right. One recognises the ring and the loss-cone.}}\label{fig-dist-2}
\vspace{-0.3cm}
\end{figure}

{As expected, they do not exhibit signatures of the short-scale electron-holes in Figure \ref{fig-electrons-eh}. The dip on the top of the green curve results fro the temporally weaker electron fluxes in slice A at maximum and cannot be taken as clear indication of electron holes. The distribution in slice B shows even less fluctuation, except for the statistically confirmed peak at its high-speed flank before exponential cut-off. This peak results from the repeated higher energy spots separated from the main distribution in slice B in Figure \ref{fig-electrons-eh}. In consultation with numerical simulations we will below argue that it might be due to the presence of electron holes.} 

{We have not attempted to determine a perpendicular distribution as no electron holes are expected to be present there anyway. Since holes cannot be resolved in the electron distributions, we have also not attempted to determine an oblique slice of the distribution function. It would not provide any deeper insight into the structure of the electron distribution.}

{As for an example of a full distribution we plot in Figure \ref{fig-dist-2} one of the best measured full electron ring-horseshoe distributions obtained under similar conditions but taken from another auroral passage. In those measurements the partial ring pops out at high parallel weakly relativistic velocities just below $5\times 10^4$ km/s. The distribution exhibits a well expressed loss cone at negative $v_\|$ while at pitch angles near $90^\circ$ the ring disappears and flattens to fill the gap to the background. This flattening is also found in the parallel direction in Figure \ref{fig-dist} where, however, it is not strong enough to completely deplete the ring. In both cases flattening and filling the gap is probably due to wave- particle interactions mostly, however, caused by VLF.  It has been shown that VLF is excited at high growth rate in the presence of positive perpendicular velocity gradients \citep{labelle2002}. Whether the electron-cyclotron horseshoe maser is involved into the depletion of the distribution remains an unanswered question. Its growth rates are smaller than VLF growth rates and, in addition, at the extremely low plasma densities in the auroral kilometric source the confinement of the free-space waves which is necessary for causing quasi-linear depletion of the ring distribution is  so far unproven. In fact, the X-mode cut-off and upper hybrid frequency coincide within few tens of Hz at plasma frequency $\sim 2$ kHz and cyclotron frequency $\sim 400$ kHz yielding a cut-off at $\sim (400+0.1)$ kHz. Excitation of the X-mode at $\sim\omega_{ce}$, which corresponds to wave lengths $\lambda\lesssim1$ km, prevents escape of the waves only for very steep plasma gradients. Such gradients exist at the boundaries of the large-scale auroral cavity where the radiation is scattered back; but for confinement multiple scatterings are required which also bring the wave out of resonance. Quasilinear calculations and simulations of the electron cyclotron maser emission assume complete confinement and resonance over the entire box and thus grossly overestimate quasilinear depletion.}

\section{Electron hole physics - A focussed review}
{In order to blame electron holes for their contribution to the generation of electron cyclotron maser radiation, electron holes require producing steep \emph{perpendicular} velocity gradients on the electron distribution function.}  

In this section we sketch the process which leads to formation of electron holes. Historically, electron holes have been predicted \citep{dupree1972,dupree1982,dupree1983} being caused due to the grainy structure of the momentum space occupied by a collisionless plasma. In such a collisionless case the particles are not densely packed in configuration space being separated by quite large spatial distances which causes electromagnetic fields to act between them over these distances. Since the action is not instantaneous but is propagated by plasma oscillations at finite phase and group velocities, the distribution of particles (electrons in this case) cannot become completely smooth yielding lumps of electrons separated by regions of relatively strong electromagnetic fields. Plasma fluid theory indeed reproduced solutions of this kind: solitons, Bernstein-Green-Kruskal modes \citep{schamel1975,schamel1979,schamel1983} and also genuine electron holes sometimes called microscopic double layers or even electrostatic shocks \citep[for an extended review of theory and experiment cf., e.g.,][]{schamel1986}. But the dynamics of such structures could only become investigated more closely with the help of numerical particle simulations. 

\begin{figure}[t!]
\centerline{{\includegraphics[width=0.475\textwidth,clip=]{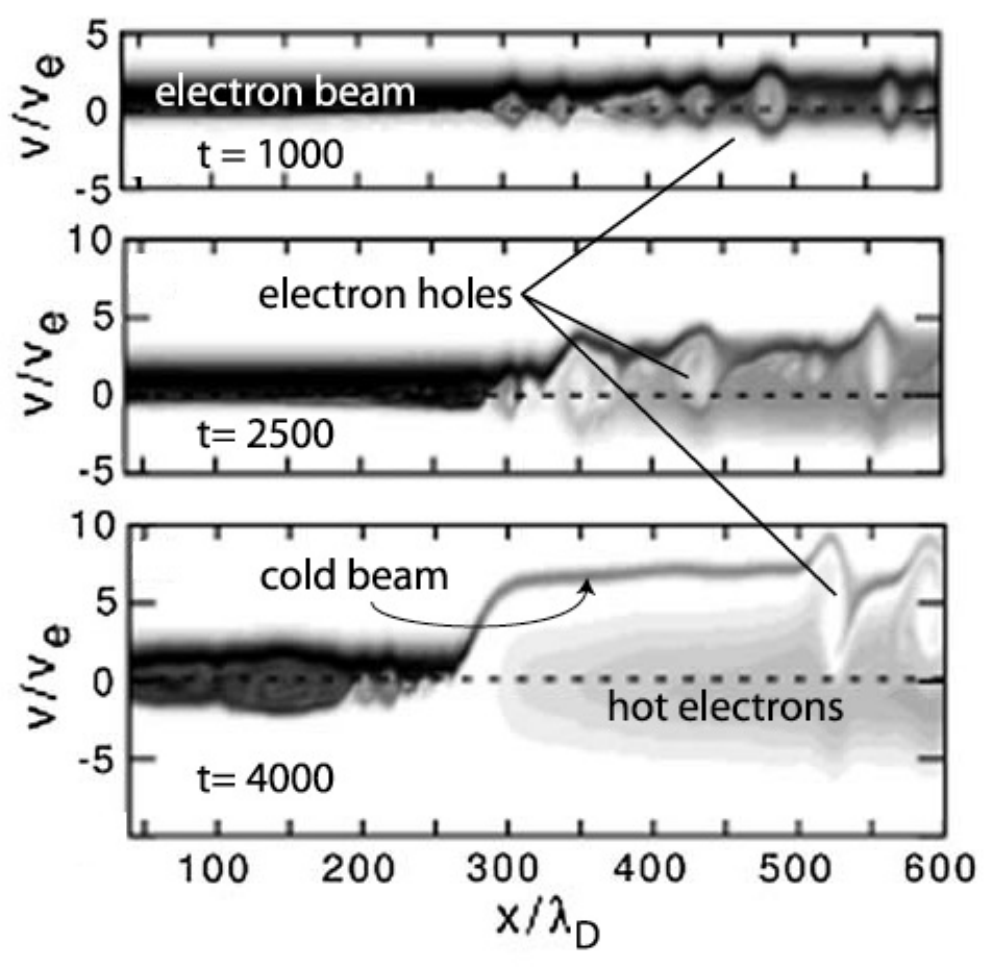}
}}
\caption[ ]
{\footnotesize The electron momentum phase space in a one-dimensional simulation of the formation of electron holes under the conditions of a Buneman unstable plasma \citep[data compiled from][]{newman2002}. These are three snapshots of the evolution of the electron distribution function corresponding to the direction parallel to the ambient magnetic field after start of the instability. Times are measured in inverse electron plasma frequencies (electron-plasma periods) $\omega_e^{-1}$, spatial scales are in Debye lengths $\lambda_D$. The initial electron beam is warm with ion-frame drift  $|V_e|$ exceeding the electron thermal speed. Ions, not shown, move in the opposite direction. At $t\omega_e=10^3$ Buneman mode evolution and propagation is seen to become localised evolving into the electron beam distribution, trapping and reflecting electrons out of the beam. Growing holes at time $t\omega_e=2500$ coalesce to form common structures, some structures becoming very extended in velocity and configuration space at final simulation time $t\omega_e=4000$. The large hole has split the initial hot beam into a very cold beam of seven times initial velocity and a very dilute hot trapped electron distribution of about ten times initial thermal spread. One should note the sharp positive velocity gradient between the holes and the cold beam and compare it to the weak negative velocity gradient of the hot trapped electron distribution. }\label{fig-SIMul}
\vspace{-0.3cm}
\end{figure}

These simulations showed a number of unforeseen effects of which the most interesting is that electron holes are genuinely non-stationary. An instructive example is reproduced in Figure \ref{fig-SIMul}  \citep[adapted with changes from][]{newman2002}. Before, however, coming to describing what physically important conclusions can be drawn from it and what their relevance is in view of the electron-cyclotron maser radiation we have in mind, it is necessary to briefly go into the mechanism by which such holes arise in a completely collisionless plasma. 

There are two stages for the evolution of holes in plasma. The first is the identification of the relevant plasma instability which amplifies the electric field choosing a particular range of wave numbers and frequencies from the always available thermal background fluctuation spectrum. The second stage refers to the nonlinear evolution of the hole and its possible long-term behaviour which might result in saturation. Simulations demonstrate that saturation is not achieved because of non-stationarity of the holes, hence this second stage is only intermediate.

The instability relevant for production of electron holes is the instability of the Buneman mode. Since, for the production of radiation, we are interested only in electron structures\footnote{Ion structures do not produce any relevant radiation if not changing the electron distribution. This might, however, be the case under some conditions of current flow when ion holes evolve, interrupt the current on  longer-time and larger spatial scales. Such effects will not be considered in this Letter.} we take the ions as immobile and restrict to electron motion and electron driven instability. It has been shown in this case that for sufficiently strong current flow along the magnetic field $\mathbf{B}$ two important instabilities evolve in the plasma, the purely electrostatic Buneman mode \citep{buneman1958,buneman1959}, and the purely electromagnetic Weibel \citep{weibel1959} or Weibel-filamentation \citep{fried1959}  modes. The former generates high-frequency $\omega_i\ll\omega_B\ll\omega_e$ electrostatic field fluctuations on scales $\lambda_B\gtrsim $ several $\lambda_D$ of the order of several Debye-lengths; the latter generates very low frequency $0\lesssim\omega_W\ll\omega_{ce}$ magnetic fluctuations on scales of the order of the electron-inertial scale $\lambda_W\sim\lambda_e=c/\omega_e$. Here, $\omega_i,\ \omega_e$ are the respective ion and electron plasma frequencies, $\lambda_D=v_e/\omega_e$ the Debye length, $v_e=\sqrt{2T_e/m_e}$ the electron thermal speed, $T _e$ electron temperature (in energy units), $m_e$  electron-rest mass, and $\omega_{ce}=eB/m_e$ the (non-relativistic) electron cyclotron frequency in the magnetic field of strength $B$. This mode is primarily unimportant in the process of generation of radiation. However, in very fast (possibly relativistic) current flows $|V_e|\gg v_e$ it may generate sufficiently strong transverse magnetic fields to deflect the electrons \citep[and even the ions, in which case it leads to formation of small-scale shock waves, cf., e.g.,][and references therein]{treumann2011} and thus modify the Buneman instability conditions.

Though both modes may be simultaneous, for electron-hole formation only the Buneman mode is of interest. It is excited when the electron-current drift velocity $|V_e|>v_e$ along the magnetic field -- where $V_e$ is measured in the ion frame, i.e. it is the effective relative average drift velocity  between the electrons and the ions leading to current flow -- exceeds the electron thermal velocity $v_e$. In this case the central wave number of the Buneman mode becomes $k_B\approx \omega_e/V_e$, and its frequency $\omega_B$ and growth rate $\mathrm{Im}(\omega_B)$ become respectively\vspace{-1mm}
\begin{equation}
\hspace{7mm}\omega_B\ =\ \left(\frac{m_e}{16m_i}\right)^{\!\!\!\!\frac{1}{3}}\hspace{1.2mm}\approx\ 0.03\ \omega_e, 
\end{equation}
\vspace{-5mm}
\begin{equation}
\mathrm{Im}(\omega_B)\ =\ \left(\frac{3}{16}\frac{m_e}{m_i}\right)^{\!\!\!\!\frac{1}{3}}\approx\ 0.05\ \omega_e
\end{equation}
indicating that the Buneman mode is a very fast growing (reactive) mode with growth rate of the order of its frequency \citep[][Chpt. 2, pp. 21-25]{treumann1997}.

The condition on the wave number indicates that the Buneman wave-length $\lambda_B=2\pi/k_B\sim V_e/\omega_e>\lambda_D$ exceeds the Debye length by the factor $V_e/v_e$. Initially, its phase velocity is easily found to be a fraction of the electron current drift speed
\begin{equation}
V_B\ \equiv\ \frac{\omega_B}{k_B}\ \sim\ 0.03\ |V_e|
\end{equation}
During the evolution of the instability and formation of electron holes this velocity increase{s due to} momentum exchange between the plasma and the hole until the hole is speeded up to nearly current flow velocity. 

The fast-growing small-scale high-frequency electric fields amplified by the Buneman instability grow on the expense of the electron current drift along the magnetic field. Those electrons which are slow, i.e. of energy $\epsilon_e<eU_B^\mathit{max}$, where $U_B^\mathit{max}\sim \frac{1}{2}E_B^\mathit{max}\lambda_B$ is the maximum (or minimum) potential drop in the Buneman mode, are scattered by the Buneman mode and, depending on the sign of the electric field become either trapped in or expelled from the region of the wave. This effect is the well-known action of the ponderomotive force exerted by a fluctuating electric potential field on the electric charges in question, this time the highly mobile electrons. 

Thus the Buneman mode causes a spatial structuring of the otherwise homogeneous plasma; in its deep negative potential wells it traps electrons whose energies are low in the proper frame of the wave, while it repels all those electrons from the regions of high positive wave potentials. In this way it also causes a spatially varying separation of the electron distribution into two components, a trapped component and an escaping accelerated component. This is the essence of the physics of Bernstein-Green-Kruskal modes and the ingredient of the model calculations of \citet{schamel1975,schamel1979} and \citet{muschietti1999a,muschietti1999b}. 

However, in addition, the numerical simulations show that this state is not final. It cannot be kept for ever simply because the trapped electrons have their own dynamics, and in addition the reaction on the current flow causes scattering of the current and thus weak dissipation which locally decreases the current velocity until it drops below the electron thermal speed respectively the electron temperature is artificially increased by the split in the distribution function and the condition for excitation of the Buneman mode becomes violated. Then the instability ceases and the holes become destroyed. These times amount to many plasma periods, times of the order of $t_\mathit{fin}\omega_e\sim10^5$ or longer.  

When this happens, the nature of the instability changes because at $|V_e|<v_e$ the ion-acoustic instability takes over and dominates the further evolution of the plasma. However, since the plasma is highly structured at these times, the ion-acoustic instability becomes different from that in a homogeneous plasma, and the strong inhomogeneity on the scale of the ion-acoustic wave length must be taken into account. This is a formidable task; but its essence is that the ion-acoustic instability leads to a further structuring of the plasma with formation of larger scale holes in which now the ions are also involved and which flow at much slower speed than the electron holes along the current staying behind the flow. 

An additional modification concerns the obliqueness of current flow with respect to the magnetic field. In an oblique current the instability changes character becoming the modified two-stream instability with typical frequency near the lower-hybrid frequency $\omega_\mathit{lh}$. The nature of hole formation changes in this case as well. The parallel component of the current will still drive the Buneman mode unstable as long as $V_{e\|}>v_{e\|}$, but the perpendicular component excites lower-hybrid modified-two stream waves. 

{Numerical  Vlasov simulations \citep{newman2001,newman2002,ergun2002} as well as model calculations \citep{muschietti1999a,muschietti1999b} indicate in addition, that electron holes are very narrow in configuration space while occupying a substantial part of parallel velocity space. Moreover, two-dimensional and three-dimensional simulations \citep[][]{oppenheim2001,newman2002} show that the structures in configuration space are unstable in the perpendicular spatial direction against bending, breaking off into smaller spatial structures and radiating VLF waves.  When artificially injecting a low-amplitude oblique electrostatic seed-whistler in order to control the interaction with the holes in the perpendicular spatial direction \citep{newman2002} it is found that bending sets in after about $t\omega_e\sim1500$ simulation times, and after roughly $t\omega_e\sim3000$ simulation times perpendicular destruction of the holes is completed leaving structures of same parallel and perpendicular spatial sizes while having amplified the seed-whistler. In this case the bending and destruction of the holes occur entirely in configuration space.}\footnote{{No electron data on the full velocity space in simulations are available, however, at least to our knowledge, even though in the electrostatic (non-magnetised) simulations perpendicular velocities are included. Also in magnetised simulations perpendicular velocity space data are not presented. The reasons are that the perpendicular velocity structure of the hole is not of interest to most simulationists if not dealing with radiation, and those who are interested in radiation consider electron holes as being irrelevant.}}

These considerations set the frame for our discussion of electron-hole generated electron-cyclotron maser radiation in the following sections. 

\section{Summary of electron hole properties}

The fact which can be learned from the simulation shown in Figure \ref{fig-SIMul} and briefly described in the caption to that figure can be summarised semi-quantitatively to the extent of their relevance for the electron-cyclotron maser under auroral-kilometric conditions as follows:

\subsection*{\small Simulation results}

~~(i)  Electron holes grow from a-few $\lambda_D$ scales to a few-10\,$\lambda_D$ scales along the magnetic field. By growing{, electron holes} in close proximity coalesce both in configuration and momentum space leaving behind a long wake of dilute hot electron plasma. This is seen from comparing the snapshots taken in the first and third panels in Figure \ref{fig-SIMul} which are separated by $\Delta t\omega_e=3000$ time steps {which corresponds to $\Delta t\sim 1$ s in the observations}.

(ii) In velocity space they grow {and move into} the region of the beam. This effect is due to momentum exchange with the beam. The holes{. lacking electrons, resemble positive charges on the negative electron background and} become attracted by the beam, an effect which has been described analytically for narrow holes \citep{treumann2008}.

(iii) Holes split the electron velocity distribution into a narrow cold accelerated-beam distribution and a broad hot though very dilute distribution. The generation of the cold beam is a very interesting {side} effect as it provides a mechanism of strong electron cooling and acceleration. The cooling amounts to roughly a factor $>10$ in temperature, the acceleration to a factor $\sim 50$ in energy.

(iv) The hot {electron distribution trapped inside the hole} observes a comparable temperature increase of factor $\sim 10-20$ in temperature. These electrons, in the ion frame, have about symmetric propagation direction, positive and negative along the field, typical for particles being electrically trapped and bouncing back and forth between the walls of the potential well. {Their} density is rather dilute {because} only a restricted number of electrons can become trapped and kept in a potential trough.

(v) For the electron cyclotron maser most important is the production of a steep positive velocity gradient between the trapped distribution inside the hole and the cold electron beam bounding it from the high speed side, and a rather weak (flat negative) velocity gradient toward the interior of the hole. The cyclotron maser mechanism requires the velocity space gradient in the perpendicular direction which in numerical simulations is not considered. {Below we justify the existence of such a perpendicular velocity space gradient.}

In order to apply the insights obtained from the simulations we need to approximately quantify them with the help of the observations shown in Figures \ref{fig-akr-spectrum} and \ref{fig-dist}. and we have to translate the one-dimensional simulation results into the {two-dimensional velocity-space plane} of the anisotropic electron {ring-horseshoe} distribution like the one observed in the auroral kilometric radiation source (i.e. Figures \ref{fig-dist} and \ref{fig-dist-2}).

\begin{figure*}[t!]
\hspace{-5mm}\centerline{\includegraphics[width=0.5\textwidth,clip=]{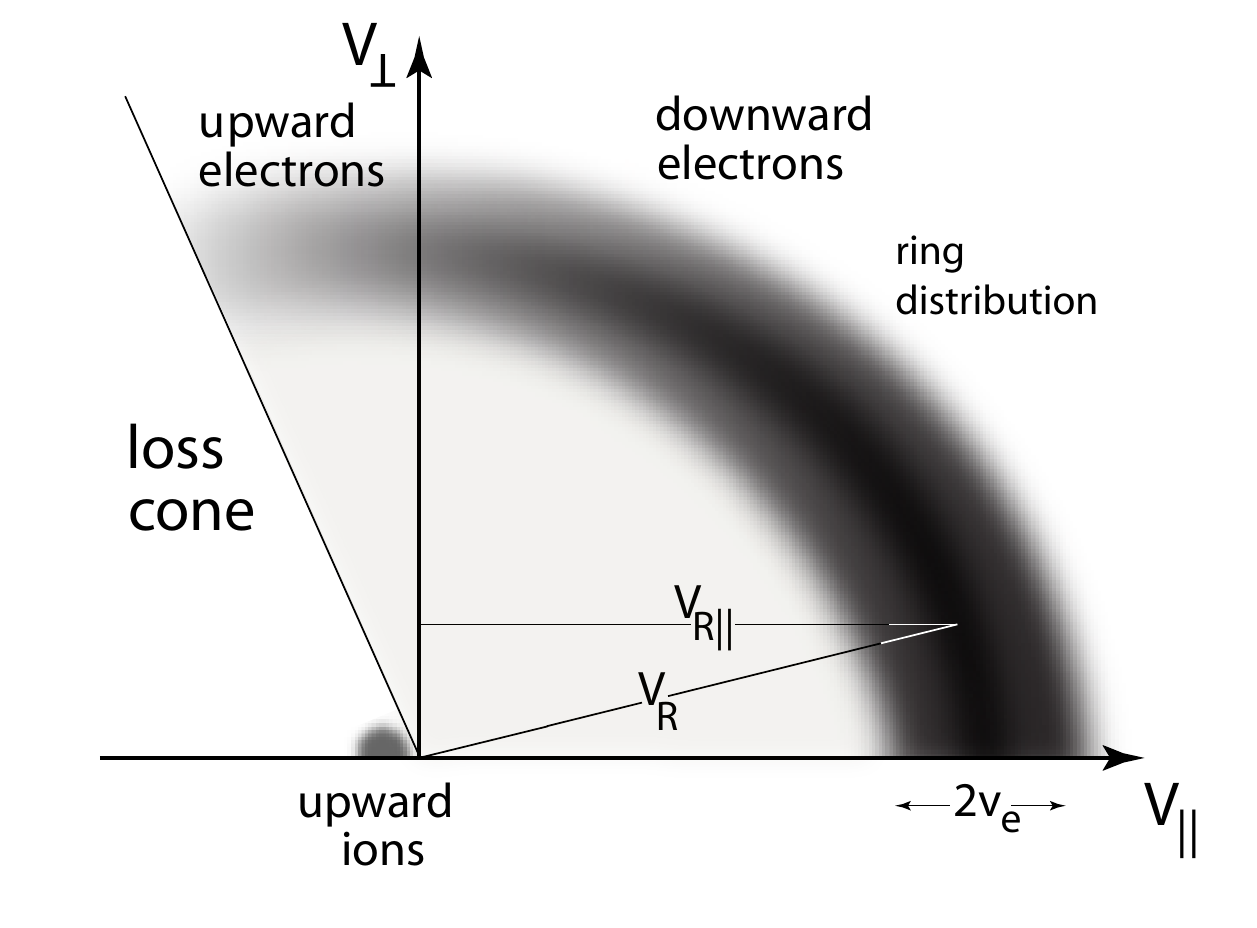}
\includegraphics[width=0.5\textwidth,clip=]{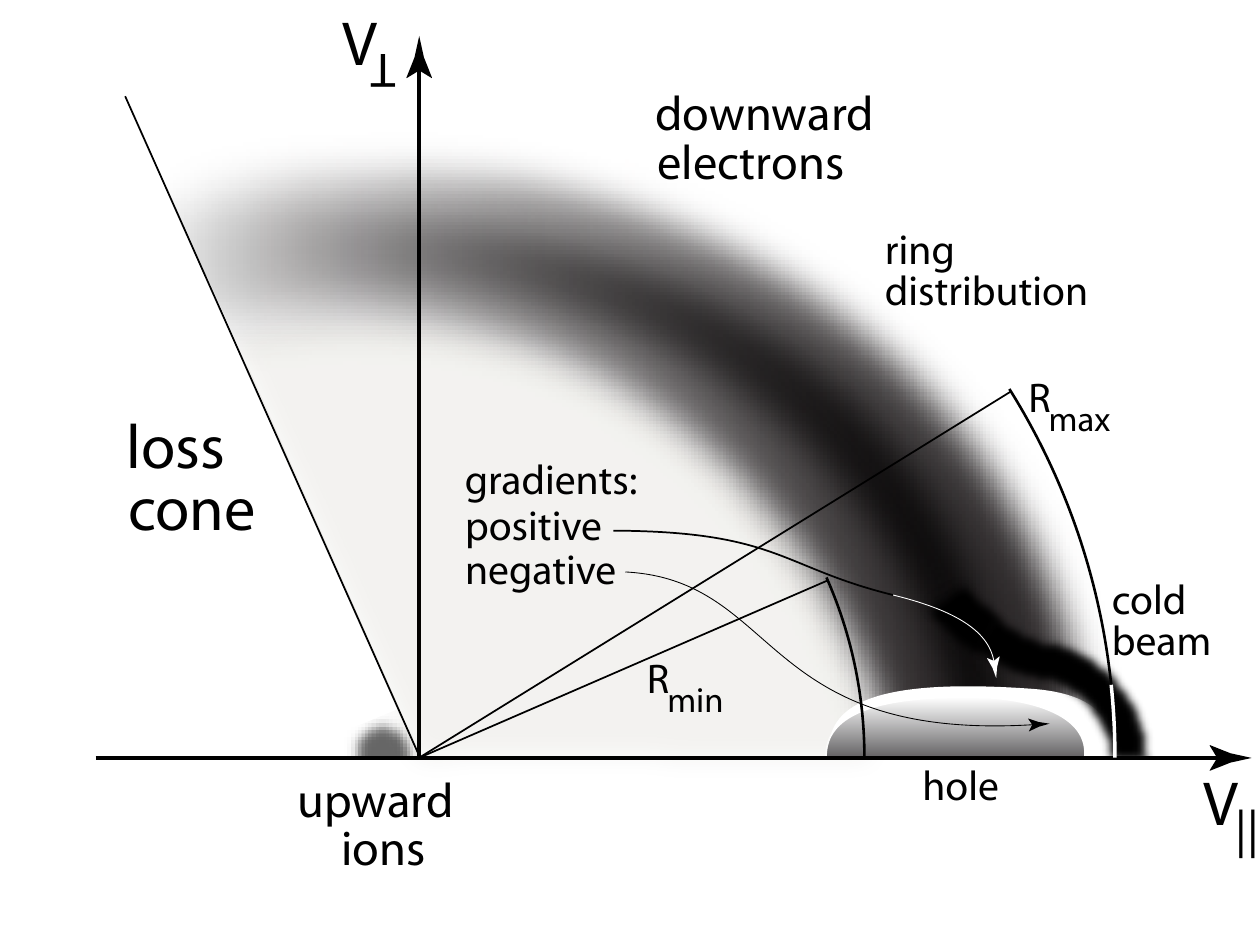}}
\caption[ ]
{\footnotesize A schematic of a broad electron hole evolving in velocity space under the condition of a loss-cone-truncated ring-(`horseshoe')-distribution (shown on the left)  with parallel velocity spread $2v_e$ substantially less than the (parallel) ring speed $V_{R\|}$. Under this condition the two-stream instability evolves and generates electron holes on the low parallel-velocity side of the warm-beam distribution (as shown on the right). The hole grows and is attracted into the beam into parallel direction. It traps electrons and accelerates part of the beam to high parallel speeds into a fast cool beam. Its finite extension in perpendicular velocity implies the production of perpendicular velocity-space gradients as shown by arrows. A steep positive gradient exists at higher perpendicular velocities pointing from the interior of the hole into the beam distribution, while a flat negative gradient points from the boundary of the hole into the warm trapped electron component which fills the hole. }\label{fig-hole-rad-1}
\vspace{-0.3cm}
\end{figure*}

\subsection*{\small Translation into observed quantities}

Quantification is done by listing the known parameters of the auroral kilometric radiation source region. All the following numbers are approximate only, though. This, however, has little effect on the main argument and conclusions. 

The plasma frequency in the range under discussion during auroral kilometric conditions is of the order {of $\omega_e/2\pi\approx1.5-2.0$ kHz} corresponding to very low plasma densities. The electron temperature, on the other hand is $T_e\sim$ few keV, or a thermal velocity of $v_e\lesssim 2 \times 10^4$ km\,s$^{-1}$,  compared to a nominal {bulk downward ring-horseshoe beam} velocity of $V_e\approx (2.5-3)\times10^4$ km\,s$^{-1}$. This corresponds to a Debye length of $\lambda_D\lesssim 1$ km and a {spatial} extension of the electron hole {parallel to the magnetic field} of the order of $L_\|\sim(1-3)$ km. The electron gyro-radius at the velocity of the bulk flow is about $r_{ce}\approx $ few times 10 m. The thermal electron gyro-radius is roughly $r_{ce}^\mathit{th}\sim 8$ m. Hence, on the scale of the hole the electrons are magnetised even when trapped. {When heated by a factor of 10--20 this becomes by a factor of 3--5 larger, i.e. 20--30 m.} 

{What concerns the transverse scale of the holes in velocity space we note that \citet{franz2000} observed about spherical holes in configuration space in the $\omega_{ce}\gg\omega_e$ auroral case which corresponds to the auroral kilometric radiation source of interest here. In \emph{weak} magnetic fields it is claimed from numerical experiments that the holes are elongated perpendicular to the magnetic field due to the large gyroradii of electrons  \citep{oppenheim2001}, forming pancakes in configuration space. This does not apply to the auroral radiation source, rather it holds in the solar wind.} 

{Electron holes thus} extend in configuration space also in perpendicular direction, at least over distances of the trapped electron gyro-radius.  {This implies that, in the frame of the moving hole, the ring-horseshoe distribution becomes substantially distorted by the presence of the hole also in perpendicular velocity up to perpendicular velocities of the trapped particles}. This is schematically shown in Figure \ref{fig-hole-rad-1} on its right-hand side. {Referring to the simulation results, the distortion of the ring distribution implies generation of a fast gyrating cold beam and trapping of lower energy electrons inside the hole. In addition, however, it also implies the production of a steep positive perpendicular gradient in velocity space over some distance in $v_\|$ and, at the same time, a weak negative gradient in perpendicular velocity pointing into the hole toward the trapped electron population.}  

It we take the acceleration of the persistent cold electron beam in the wake of the merged electron holes in Figure \ref{fig-SIMul} to be roughly $5v_e$, then its velocity becomes $V_b\lesssim10^5$ km\,s$^{-1}$. This is already truly relativistic. The temperature of this cold beam on the other hand is substantially less, of the order of $T_b\lesssim 1$ keV. The hot trapped electrons have instead temperature of the order of $\sim 5$ keV comparable to the thermal spread of the ring distribution.  

{Unfortunately little is known about the relative phase space densities from the simulations; acknowledging, in addition, that the simulations demonstrate the evolution of electron holes but are not representative in comparison to observational number densities, we may refer to Figure \ref{fig-dist} instead.} The little high-speed peak at the curve labelled `high', {when interpreted as an averaged-over signature of the cold accelerated beam,} suggests that the phase space density of the cold beam is roughly $10^{-2}$ times that of the {auroral electron beam forming the global ring-horseshoe distribution.} The density is still low, dominated, however, by the horseshoe (the background being mainly photo-electrons). {Hence, the phase space density of the relativistic cold beam becomes about $N_b\approx 2\times10^{-3} $ cm$^{-3}$. The density of the trapped particles in the centre of the hole is not known since the hole is not resolved by the instrument. It is probably another factor of $10^{-2}$ or even more less.  However, in estimating the effect on the radiation we need only relative numbers, as we will argue in the next section}.

\subsection*{\small{Transverse extension of holes in phase space}}
{As promised, this subsection provides an argument for the limited perpendicular extension of electron holes in velocity space.  The above reasoning implies that trapped electrons are readily thermalised. In this case the hole extends up to trapped thermal velocities $v_\perp\lesssim v_{e\,tr}\sim 2.5\times10^4$ km/s in $v_\perp$, i.e. almost up to angles of 60$^\circ$ in Figure \ref{fig-dist-2}. }

{A more physical argument going beyond the simple previous argument based on the electron gyroradius can be provided as follows when referring to the above reviewed theory that electron holes are generated by the Buneman instability. This instability requires that the \emph{parallel} electron (current) drift velocity in the ion frame must exceed the thermal spread of the electron distribution. The ion speed is about $V_i<10^3$ km/s directed upward along the magnetic field (which would add to the electron speed) but which can be taken as zero in the auroral case. For gyrating electrons with perpendicular velocity $v_\perp$ this condition should hold for each fixed $v_\perp=$ const. }

{In a ring distribution we have $v_\perp=V_R\sin\theta$ with $\theta$ the pitch angle. Hence, at
where we used the observational values for $v_e$ and $V_R$. These numbers correspond to pitch angles of $\theta\lesssim 30^\circ$ up to which the Buneman instability is unstable at finite $v_\perp$, setting a limitation on the perpendicular extension of electron holes in velocity space. (One should note that electrons are magnetised and therefore there is only one perpendicular velocity coordinate, not two.) This justifies our assumption that any electron holes generated by the Buneman instability will necessarily have a finite extension in perpendicular velocity space. Because of this extension any electron holes in magnetised plasmas produce steep velocity space gradients on the electron distribution function which are restricted to the edges of the holes in parallel and perpendicular direction.}

{The above expression may be used to estimate the shape of an electron hole in phase space. Dividing by $v_e$ we obtain $v_\perp/v_e<0.5V_R/v_e\approx 0.5-0.7$. Since the parallel extension of the hole is $\Delta v_\|< v_e$ the two axes of the hole are approximately comparable, at least within the uncertainty of our measurements. In a first approximation it will thus not be unreasonable to assume a circular cross section shape of the hole in particular as the observations of \citet{franz2000} render some experimental support to this assumption.}

\section{Cyclotron-maser radiation from single electron holes}
Inspection of the schematic in Figure \ref{fig-hole-rad-1} suggests that any cyclotron-maser radiation that is related to the electron hole should exhibit a band of emission and a band of absorption which will appear in tandem. The emission band is caused by the steep positive velocity gradient at the high-speed boundary of the hole; conversely the absorption band is caused by the negative velocity gradient pointing from the hole boundary into the centre of the hole. {Since} both gradients occur at different places {in velocity space}, a gap in frequency space develops between them which is caused by the difference in the resonance conditions. This fact complicates the theory with respect to the consideration of the simple ring distribution or the Dory-Guest-Harris distribution \citep[for the latter case see][]{pritchett1986}. 

However, there is another serious complication which is introduced by the geometric form of the hole in velocity space. The electron hole in Figure \ref{fig-hole-rad-1} is not centred at the origin of the velocity space. Hence, in principle, the resonance condition should become a displaced ellipse in velocity space as known from the general resonance. The implication is that electron holes are well suited to 

(i) generate oblique electron-cyclotron maser radiation emitted at a finite angle against the magnetic field; 

(ii) this radiation will for the same reason in general be in a frequency range that is not restricted to the immediate vicinity of the electron cyclotron frequency but can occur at higher frequencies quite above the electron-cyclotron frequency.

\subsection*{\small Purely perpendicular radiation - plasma frame approach}
Here, however, we will, for a first attempt and demonstration that electron holes are effective radiators, restrict to the conventional consideration of just purely transverse emission with wave number ${\bf k}=k_\perp \mathbf{\hat e}_\perp,~k_\|=0$,  where $\mathbf{\hat e}_\perp$ is the  unit vector perpendicular to the magnetic field $\mathbf{B}$. In this case it is well known that the resonance condition of the electron-cyclotron maser instability simplifies substantially becoming, with $u=\gamma\beta$ the normalised to $c$ 4-velocity of the electrons, $\beta=v/c$, and $\gamma=(1-\beta^2)^{-\frac{1}{2}}=(1+u^2)^\frac{1}{2}$ the Lorentz factor,\vspace{-1mm}
\begin{equation}\label{eq-res}
u_\perp^2+u_\|^2 -2(1-\nu_{ce})=0
\end{equation}
where $\nu_{ce}=\omega/\omega_{ce}$ is the non-relativistic ratio of wave frequency to electron cyclotron frequency. Clearly, resonance is possible only for $\nu_{ce}\lesssim 1$ and, as usual, the resonance line is a circle in the normalised 4-velocity plane $(u_\perp,u_\|)$ of radius 
\begin{equation}
R_\mathrm{res}=\sqrt{2(1-\nu_{ce})}
\end{equation}
located between the minimum and maximum radii $R_{min}, R_{max}$ of resonance which correspond to maximum and minimum resonant frequencies, respectively.  
 
The relativistic cold plasma dispersion relation of the R-X mode,  which is the real part of
\begin{equation}
n^2-1-(2-n_\perp^2)A=0, \quad\mathrm{with}\quad n^2=k^2c^2/\omega^2
\end{equation}
allows for a range of such resonant frequencies below the non-relativistic $\omega_{ce}$ which is quite narrow \citep[cf., e.g.,][]{pritchett1986}, depending on the plasma parameters. {The function $A$ (defined below) contains the plasma response.} The maximum resonance frequency (minimum radius) is very close to $\nu_{ce}=1$. On the other hand, the minimum resonance frequency can be estimated from the cold X-mode dispersion relation yielding 
\begin{equation}
\nu_{ce, min}\approx 1- \frac{\gamma^2}{2}\left(\frac{\omega_e}{\omega_{ce}}\right)^2
\end{equation}
Taking $k_\|=0,\ |1-\nu_{ce}|\ll 1$, one has to first order $\mathrm{Re}(A)\ll 1$. This expression is to be used in the calculation of the growth rate 
\begin{equation}
\mathrm{Im}(\nu_{ce}) \simeq-{\scriptstyle\frac{1}{2}}\mathrm{Im}(A)
\end{equation}
Maser radiation will be produced if $\mathrm{Im}(A)<0$ is negative (corresponding to `negative absorption' of the electromagnetic R-X mode). The resonant imaginary part of $A$ is given by \vspace{-3mm}
\begin{eqnarray}\label{Im-A}
\mathrm{Im}(A)=&-&\frac{\pi^2}{2\nu_{ce}}\frac{\omega_e^2}{\omega_{ce}^2}\int\limits_{-\infty}^\infty\int\limits_{0}^\infty{\rm d}u_\|u_\perp{\rm d}u_\perp^{2}\frac{\partial F_e}{\partial u_\perp}\times\nonumber \\
&\times&\delta(u_\|^2+u_\perp^2-R_\mathit{res}^2)
\end{eqnarray}

This integral must be solved accounting for resonance along the resonant circle. It is clear that only the part of the resonant circle contributes to emission which passes the positive perpendicular velocity space gradient. The remaining part along the resonant circle contributes to absorption. In the calculation we have two options. Either we draw a set of resonant circles which cross the positive gradient and for them calculate the emission in pieces over the range of frequencies selected by the resonance condition, or we transform to the phase space coordinates centred on the electron hole. In the first case we can simplify the calculation by switching to an infinitesimal approach which simplifies the analysis. In the other case the transform is along $v_\|$ and, assuming a nearly circular structure of the hole, the calculation is more general. However, the transform of the results back to the stationary system is substantially more complex. In addition, the full integral must be solved knowing the precise form of the distribution function. 
\begin{figure}[t!]
\centerline{{\includegraphics[width=0.475\textwidth,clip=]{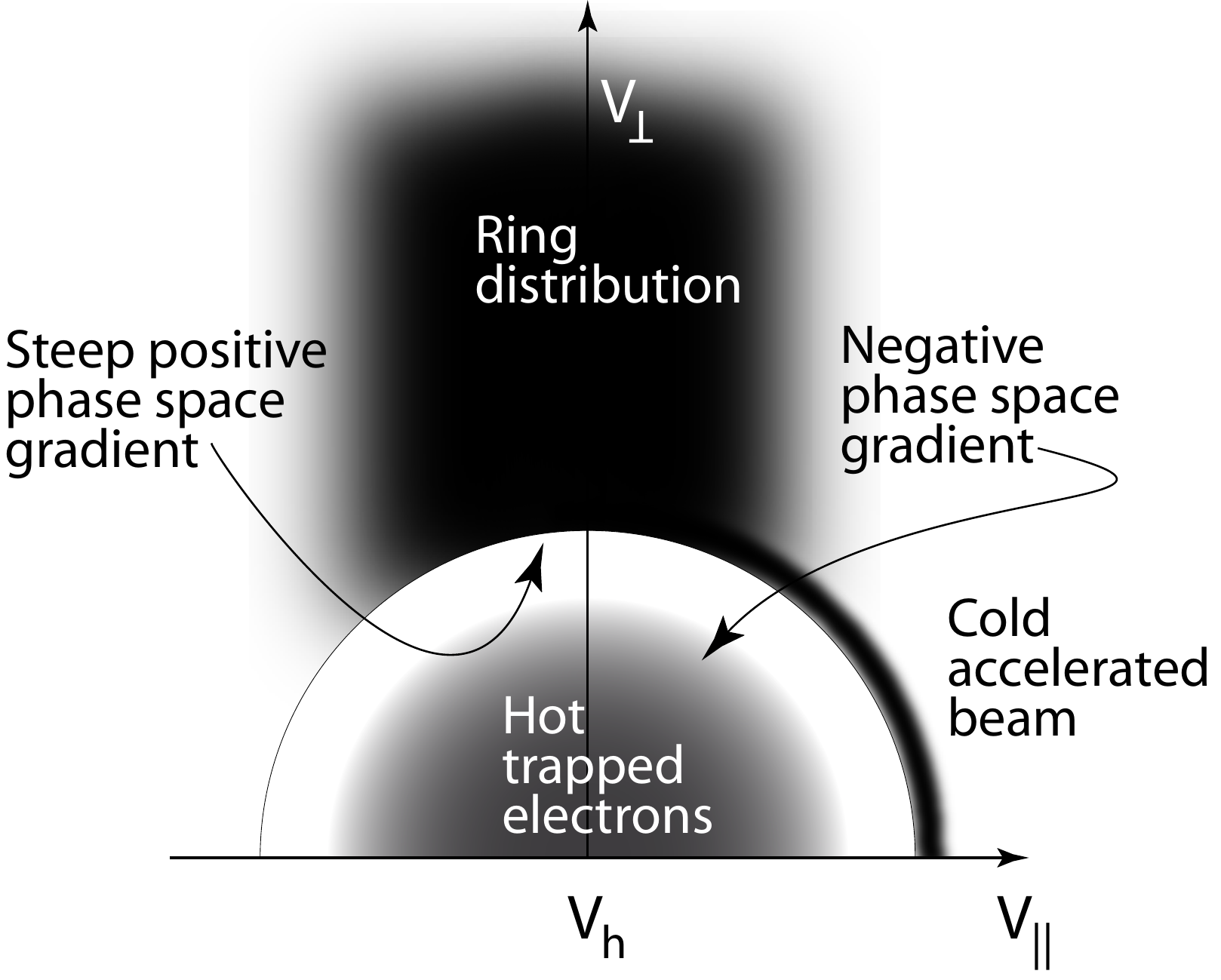}
}}
\caption[ ]
{\footnotesize The idealised circular hole model as seen in the hole-centred reference frame in momentum space as has been translated from the simulations in Figure \ref{fig-SIMul} into an instantaneous snapshot of one well developed hole. In the stationary frame the hole centre is at parallel hole velocity $V_h\sim V_R$. The fast cold beam that arises at the high velocity side of the hole is indicated. Also indicated is the hot trapped distribution. The hole has pushed the ring distribution away from its location. Phase-space velocity gradients evolve along the hole boundaries. The outer boundary of the hole shows a steep positive gradient, while the interior of the hole has a weak negative velocity gradient. Translated into the electron cyclotron maser mechanism this implies a strong emission of radiation into perpendicular direction in the hole frame from the positive gradient located near the electron cyclotron frequency, and an adjacent absorption band located at higher frequencies.}\label{fig-hole-geom}
\vspace{-0.3cm}
\end{figure}

Drawing resonance circles around the origin in Figure \ref{fig-hole-rad-1} on the right which cut through the positive velocity gradient of the electron hole the former method tells us that there is indeed a limited range $R_\mathit{min}<R_\mathit{res}<R_\mathit{max}$ which contributes to emission in perpendicular direction. This range corresponds to a limited emitted frequency range of bandwidth
\begin{equation}
\Delta\nu_{ce}\equiv\nu_\mathit{max}-\nu_\mathit{min}={\scriptstyle\frac{1}{2}}\big(R_\mathit{max}^2-R_\mathit{min}^2 \big)
\end{equation}
From Figure \ref{fig-SIMul} we read that the range of resonant radii corresponds to a range in velocity of $\Delta u_\|\approx (2-3) v_e/c=R_\mathit{max}-R_\mathit{min}\equiv\Delta R_\mathit{res}$. With $\beta_e=v_e/c$ we thus have approximately 
\begin{equation}
R_\mathit{max}^2-R_\mathit{min}^2\approx 2R_\mathit{max}\Delta R_\mathit{res}\approx 2\times (2-3)\beta_e^2
\end{equation}
which yields a relatively narrow bandwidth of 
\begin{equation}\label{eq-bw}
\Delta\nu_{ce}\approx (2-3)\beta_e^2 
\end{equation}
for the radiation emitted into perpendicular direction that is centred around the maximum emitted frequency. At nominal thermal energy of $T_e\sim 1$ keV this bandwidth corresponds to $\Delta\nu_{ce}\approx 0.002$. Taking the frequency of auroral kilometric radiation to be $\omega/2\pi\sim\omega_{ce}/2\pi\approx (300-400)$ kHz, the bandwidth caused by the extension of the hole should amount to $\Delta\omega/2\pi\approx 1$ kHz only.  Though this is a narrow bandwidth, it is in the range of observation (see Figures \ref{fig-akr-spectrum} and \ref{fig-rad}) and thus is of substantial interest when comparing with observations because there is no reliable theory available so far which could explain how the extremely narrow bandwidth of the fine structure is generated in the auroral kilometric radiation (and in other emissions in space, like solar type IV bursts, which are attributed to the electron cyclotron maser).

A similar bandwidth is obtained for the absorption which, however, is at a different frequency because the range of resonance radii cut through the negative velocity gradient at lower perpendicular velocities. This implies from the resonance condition Eq. (\ref{eq-res}) that 
\begin{equation}
\nu_\mathit{abs}>\nu_\mathit{rad}
\end{equation}
The absorption at perpendicular emission is found at higher frequency than the emission.

\subsection*{\small{Working in the hole frame}}

So far we dealt just with perpendicular emission in the cyclotron maser theory not having shown yet that positive growth rates are really produced, though their existence should be clear from simple intuition. However, this approach is, as has been noted above, not entirely appropriate for emission from a small-scale structure like a hole. In order to be more general in the calculation of the growth rate, we switch to the second approach by shifting the origin of velocity space into the very centre of the electron hole. This can be safely done without any restriction because the emission from the main ring distribution in this case plays no role and can be ignored for our purposes, which are to investigate what the contribution of the hole to the electron cyclotron maser would be. 

The origin of velocity space is then located at $v_\|=V_h, v_\perp=0$, shifted along the parallel velocity axis by $V_h$, the instantaneous central velocity of the electron hole which, as the simulations demonstrate (see Figure \ref{fig-SIMul}) is non-stationary on times-scales which we assume are long against the growth of the fast growing electron-cyclotron maser instability. The entire not negligible difficulty of transformation, which must be done fully relativistic, is thus delegated to the final result after having performed the calculation of the growth rate by solving for the integral in Eq. (\ref{Im-A}).

In addition let us assume that, for simplicity, the hole is of circular shape in momentum space. The hole will actually be of more elliptical shape, extended in one direction which depends on the hole formation mechanism and the ratio of hole radius to electron gyro-radius. This difference will be ignored in the following as it introduces severe and unnecessary mathematical complications not leading to further overwhelmingly important physical insight. {We will return to this question in the discussion.} Figure \ref{fig-hole-geom} shows the relevant features which must be accounted for in the calculation of the growth rate.

In this frame and with the circular shape of the hole the resonance curve is clearly of circular form. Hence, from the previous discussion, maser emission in the hole frame will predominantly be generated in perpendicular direction, and we can apply the strictly perpendicular emission theory with $k_\|=0,\ k_\perp\neq 0$ in the hole frame. Moreover, because of the assumed circular shape of the resonance, it is convenient to transform to polar velocity coordinates $0\leq v<\infty,\ 0\leq\phi\leq\pi$ using $v_\perp=v\,\sin\,\phi,\ v_\|=v\,\cos\,\phi$, phase-space volume element $v\ \mathrm{d}v\ \mathrm{d}\phi$ and 
\begin{equation}
\frac{\partial}{\partial v_\perp}=\sin\,\phi\frac{\partial}{\partial v} +\frac{\cos\,\phi}{v}\frac{\partial}{\partial\phi}, \qquad u^2=R^2_\mathit{res}
\end{equation}
with the resonance condition degenerating to the second of these equations. Figure \ref{fig-hole-geom} can be consulted in modelling the velocity and angular dependence of the particle distribution function $F_e(u,\phi)$.  Figure \ref{fig-hole-geom} suggests that there is an angular dependence of the distribution along the resonant circle at least in the positive gradient region. We can, however, stay on the safe side by assuming that the distribution along the cold beam boundary is (about) independent of $\phi$ in the range $0\leq\phi\leq\frac{\pi}{2}$ and neglecting the part of the resonant circle in the successive interval $\frac{\pi}{2}<\phi\leq\pi$. Eq. (\ref{Im-A}) for the growth rate thus becomes\vspace{-3mm}
\begin{eqnarray}\label{eq-growthrate}
\mathrm{Im}(A)=&-&\frac{\pi^2}{2\nu_{ce}}\frac{\omega_e^2}{\omega_{ce}^2}\int\limits_{0}^\infty\int\limits_{0}^\frac{\pi}{2} u\ {\rm d}u^2\ \sin\,\phi\ \mathrm{d}\phi\ \delta(u^2-R_\mathit{res}^2)\times\nonumber \\
&\times&\left(\sin\,\phi\ \frac{\partial F_e}{\partial u}+\frac{\cos\,\phi}{u}\frac{\partial F_e}{\partial\,\phi}\right)
\end{eqnarray}
The growth rate obtained from this expression will be less than the real growth rate thus giving a rather conservative estimate. On the other hand, because of symmetry we can take the distribution along the negative-gradient resonances for the absorption independent of $\phi$ over the entire interval which yields a realistic estimate of the absorption without underestimating it drastically. Furthermore, Figure \ref{fig-dist} suggests that the distribution functions of the cold beam, as well as that of the ring, can be modelled as Maxwellians in velocity with appropriately chosen thermal velocity spreads taken from the simulations in Figure \ref{fig-SIMul}. 

\subsection*{\small Maximum growth rate}
Since we are working in the proper frame of the hole the necessity to account for the relativistic modification of the velocity distributions is weak. We, therefore, boldly assume that the distributions are non-relativistic. 

With these preliminaries in mind, the calculation of the growth rate along the positive-gradient resonance can be done in the above angular interval by dropping the partial derivative with respect to $\phi$ in the expression in parentheses in Eq. (\ref{eq-growthrate}) and setting\vspace{-1mm} 
\begin{equation}
\frac{\partial F_b}{\partial u}=-\frac{u-U}{(\Delta u_e)^2}F_b, \quad F_b(u,\phi)\propto \exp\left[-\frac{(u-U)^2}{2(\Delta u_e)^2}\right]
\end{equation}
where we have replaced $F_e\to F_b$ with the beam distribution, $U=V_b/c$ is the normalised beam velocity in the hole frame, and $\Delta u_e$ its normalised thermal spread. 

From the simulations we have roughly $V_b\approx 5v_e$ and thus $U\approx 0.1$. For the density of the beam we assume from Figure \ref{fig-dist} that $N_b\sim 10^{-2} N$. Hence, since we have already extracted the plasma density from the distribution function in Eq. (\ref{eq-growthrate}), the integral is to be multiplied by the ratio of the beam-to-ring densities $\alpha\equiv F_b/F_e\propto N_b/N_e$. (Note that the ring temperature is not entering along the resonance circle that is singled out by the Dirac-function.) Performing the calculation yields
\begin{equation}\label{eq-growth}
\mathrm{Im}(A)=\frac{\alpha\pi^3}{8\nu_{ce}}\frac{\omega_e^2}{\omega_{ce}^2}\left\{\!\!\frac{u(u-U)}{(\Delta u_e)^2}\exp\left[-\frac{(u-U)^2}{2(\Delta u_e)^2}\right]\!\!\right\}_{u=R_\mathit{res}}
\end{equation}
an expression which, as expected, is negative for $u\equiv R_\mathit{res}\lesssim U$ thus yielding growth of the electron-cyclotron maser instability and emission of electromagnetic waves in the X-mode. Since maximum growth for small beam spread $\Delta u_e$ is obtained at \vspace{-1mm}
\begin{equation}
R_\mathit{res}^m\approx U-\frac{(\Delta u)^2}{U}
\end{equation}
which is close to the beam velocity $U$, the exponential becomes unity. With this expression the maximum growth rate becomes \vspace{-2mm}
\begin{equation}\label{gr}
\mathrm{Im}(A)_\mathit{max}\approx -\frac{\alpha\pi^3}{8\nu_{ce}^m}\frac{\omega_e^2}{\omega_{ce}^2} 
\end{equation}
independent of the beam velocity (in the hole frame) and beam spread. As expected, it decreases with the maximum unstable frequency $\nu_{ce}^m\equiv \omega_\mathit{max}/\omega_{ce}$, while the bandwidth of the emission is of the same order as estimated above, Eq. (\ref{eq-bw}).  The maximum unstable frequency $\nu_{ce}^m$, which is the centre frequency of the radiation band, is obtained from $R_\mathit{res}^m$: \vspace{-0.5mm}
\begin{equation}
\nu_{ce}^m\approx 1-{\frac{1}{2}}U^2\left[1-\frac{(\Delta u_e)^2}{U^2}\right]
\end{equation}\vspace{-0.5mm}

{For a numerical estimate we take the following rather conservative values $\omega_e/2\pi=2\ \mathrm{kHz}, \ \omega_{ce}/2\pi=400\ \mathrm{kHz}$ and $\nu_{ce}^m\approx 1$. The latter value is a rough approximation only holding in the hole frame for sufficiently small $U=V_b/c$. For larger $V_b\sim c/3$ or so, $\nu_{ce}^m\approx 0.95$ which, for a nominal electron-cyclotron frequency of $\omega_{ce}/2\pi=400$ kHz, is at $\omega^m\approx 380$ kHz.  With these numbers the maximum growth rate is estimated as
\begin{equation}
 \mathrm{Im}(\nu_{ce}^m)\sim \alpha \times 10^{-4}
\end{equation}
The value of the growth rate depends on the unknown density ratio $\alpha$. Assuming that $0.01<\alpha< 0.1$ the growth rate of the electron cyclotron maser emission of a single electron hole is at most moderate being of the order of $10^{-6}<\mathrm{Im}(\nu_{ce}^m)<10^{-5}$. Even though this estimate is conservative setting a lower limit on the growth rate which increases with increasing beam speed and density, \emph{the growth rate is small}. The reason is of course that even though the perpendicular gradient at the hole boundary is very steep, the size of the hole is small and thus the number of particles contributing to resonance is small such that not much energy per unit time can be fed into the wave. Nevertheless, the growth rate is not negligibly small. Its smallness is the result of the very low plasma density. The main factor which determines the order of the growth rate is the ratio of plasma-to-cyclotron frequency $\omega_e/\omega_{ce}$ in Eq. (\ref{gr}). This ratio in our case has been taken as $2/400=5\times 10^{-3}$. Plasmas of higher density provide more particles in the hole-ring boundary for supporting radiation.} 

{In a plasma of density just three times larger the growth rate becomes comparable within one order of magnitude with growth rates which have been estimated for the growth of auroral kilometric radiation accounting for the entire ring distribution.  In that case the growth rate provided by a hole of such small spatial size would not be unreasonably low. Moreover, it should be noted that those calculation from the global ring-horseshoe distribution assume a full electron ring of same electron number at all angles. The observations of Figures  \ref{fig-electrons-eh} and \ref{fig-dist-2} show at the contrary that at the most important pitch angles $\theta\sim \pi/2$ of the ring which should contribute most to the electron cyclotron maser the electron densities are lower by at least one order of magnitude. Moreover, the perpendicular phase space gradient is about completely depleted. Hence global ring-distributions should have substantially smaller growth rates. The flattening of the perpendicular velocity gradient has been discussed above. It remains unproven that it is caused by the ring maser; rather it is caused by VLF excited by the same distribution.} 

{Figure \ref{fig-electrons-eh} suggests that a hole crossed the spacecraft every 5 ms implying that 200 holes per second or $\sim$12000 holes in the minute of observation in Figure \ref{fig-over} have crossed the linear path of the spacecraft in the auroral radiation source each of them radiating though on a probably different frequency thus building up an auroral kilometric radiation band of some total bandwidth which results from adding up all contributions of all holes to the radiation. Since, however, the frequencies do not differ very much as according to the theory they are all very close to $\omega_{ce}$ the emissions will overlap and their instantaneous bandwidth and intensity will depend on the contribution of a single hole, its emission and absorption, bandwidth, location, the angle of radiation with respect to the observer (spacecraft) and -- as we will show below -- on its velocity causing Doppler shift of its emission frequency.} 

\subsection*{\small Bandwidth and radiation trapping}
{The bandwidth of emission can be found by plotting the growth rate Eq. (\ref{eq-growth}) as function of frequency $\omega$. This requires expressing the 4-velocity $u$ through $\nu_{ce}$ via the resonance condition. We may, however, take a different and even more fundamental approach realising that the bandwidth $\Delta\omega$ is determined through the real e-folding time $\Delta t_e\gtrsim [\langle\mathrm{Im}(\omega)\rangle]^{-1}$ via the uncertainty relation $\Delta\omega\Delta t_e\sim 1$.  Experimentally this implies that measuring the instantaneous bandwidth of the emission yields an upper limit on the real growth rate. Applying this reasoning to the data in Figure \ref{fig-akr-spectrum} where the bandwidth of emission can be determined quite precisely in the turning point of the emission, we obtain $\Delta\omega\approx 2\pi\times1.5$ kHz at frequency $\omega\approx 2\pi\times 440$ kHz. This yields a fairly large upper limit  of $\langle\mathrm{Im}(\omega)\rangle< 3\times10^{-3}\omega_{ce}$ for the growth rate in this case.  For {\small FAST 1773}, the time resolution does not allow to determine the instantaneous bandwidth. Taking the full bandwidth of $\sim3$ kHz yields an upper limit  $\langle\mathrm{Im}(\omega)\rangle\sim 6\times10^{-3}\omega_{ci}$, which is orders of magnitude larger than the calculated maximum growth rate $\mathrm{Im}(\nu_{ce}^m)$.}

{This obvious discrepancy between the cyclotron-maser growth rates and the above limits, in particular for {\small FAST 1773}, is quite disturbing. One possible way out of the dilemma could be provided by the amplification length. In free space this length would have to be huge in order to provide sufficient e-foldings in order to obtain an average growth rate of the above order. However, the radiation produced by the electron hole does not propagate in free space. Its wavelength is of the order of the linear scale of the hole or somewhat less. As it is generated inside the hole in the perpendicular velocity space gradient at frequency below the X-mode cut-off, it is trapped in the hole being unable to escape. It bounces back and forth between the high density walls of the hole and moves with the hole along the ambient magnetic field. Trapping inside the hole implies that it is confined to the perpendicular velocity space gradient for the entire trapping time thereby staying in resonance and being continuously amplified.}

{For a quantitative estimate of the amplification we need to know the trapping time. This can be approximated with the lifetime of electron holes against instability and destruction. From simulations this time is somewhere near $t\omega_e\sim3000$ \citep[e.g., as found by][]{newman2002}, corresponding to $t\omega_{ce}\sim7.2\times10^5$ at {\small FAST 1773}. Taking $\alpha\sim10^{-2}$, this yields a maximum theoretical growth rate of $\mathrm{Im}(\omega^m)\sim10^{-6}\omega_{ce}$, which corresponds to $\sim 2\times10^5$ cyclotron periods for one e-folding or $\sim5$ e-foldings during the life and trapping times. The wave amplitude and power may thus grow by a factor of $\sim10^2$ and $\sim10^4$, respectively. Highest measured average power spectral densities were in the range of $4\pi{\cal P}/c\epsilon_0(\Delta f)\sim3\times10^{-8}$ V$^2$/m$^2$Hz at $\omega_{ce}/2\pi\approx$ 483 kHz. Taking a bandwith of $\Delta f\approx2$ kHz this yields a maximum power of ${\cal P}\approx 8.5\times10^{-8}$ W. This implies that the background power from which the electron-cyclotron maser instability  started was in the range of some $10^{-12}$ W, which is not too unrealistic as it corresponds about to the blue background level in Figure \ref{fig-rad}. Nevertheless, these numbers are vague in particular, as the growth rates remain very small. In addition, it remains unclear how the radiation can escape from the hole to free space. Well below the X-mode cut-off of the ambient (horseshoe) plasma after decay of the hole, escape is inhibited. A further problem is caused by the motion of the hole along the magnetic field, causing changes in the local cyclotron frequency and thus modifies the resonance thereby substantially reducing the possible amplification factor.}

\subsection*{\small Absorption rate}
Before investigating the effects which can be seen after transforming back to the stationary observer {(spacecraft)} frame, we briefly investigate the rate of absorption in the hole. 

As already noted, absorption takes place in the interior of the hole, caused by the low energy (in the hole frame) though high temperature electrons whenever the hole passes through a background  bath of radiation with appropriate frequency. This may happen when it travels across the radiation field of the global ring-horseshoe distribution or when radiation emitted by other holes passes the hole. We  have noted already that the bandwidth of this radiation band will be approximately of the same order as that of the emission band. However, the central frequency of the absorption band lies at higher frequency than that of the emission, which is a consequence of the resonance condition. Still, in the hole frame for strictly perpendicular waves the frequency should lie as well below the electron-cyclotron frequency and, thus, the emission and absorption bands may mutually overlap for their finite bandwidths. 

In order to calculate the absorption rate we return to Eq. (\ref{eq-growthrate}) which in the hole frame differs only in two points: The distribution function used will be a non-shifted Maxwellian $F_h(u)$ of spread $\Delta u_h$, not a beam, and the upper limit on the angular integral becomes $\pi$ instead of $\frac{1}{2}\pi$. Moreover, for symmetry conditions the distribution function of the trapped electrons in the spherical model does not depend on $\phi$. Hence, the derivative with respect to $\phi$ vanishes identically. With $F_h(u)$ a Maxwellian, the derivative with respect to $u$ becomes 
\begin{equation}
\frac{\partial F_h}{\partial u}= -\frac{u}{(\Delta u_h)^2}F_h(u)
\end{equation}
from which it is seen that the sign of Im($A$) is positive and thus Im$(\nu_{ce})<0$ in this case; this is the condition for absorption. In order for lying inside the trapped electron distribution, the hole-frame radius of absorption-resonance is rather small, $R_\mathit{res}<\Delta u_h$, which for the absorbed frequencies $\nu_\mathit{abs}$ yields 
\begin{equation}
\nu_\mathit{abs}>1-(\Delta u_h)^2
\end{equation}
Solving for the integral yields for the absorption rate
\begin{eqnarray}
\mathrm{Im}(A_h)&=&\frac{{\tilde\alpha}\pi^3}{4\nu_{a}}\frac{\omega_e^2}{\omega_{ce}^2}\frac{1-\nu_\mathit{abs}}{(\Delta u_h)^2}\exp\left[-\frac{(1-\nu_\mathit{abs})^2}{2(\Delta u_h)^2}\right] \cr
&<&\frac{{\tilde\alpha}}{4}\frac{\pi^3}{1-(\Delta u_h)^2}\frac{\omega_e^2}{\omega_{ce}^2}\exp\left(-\scriptstyle{\frac{1}{2}}\right) 
\end{eqnarray}
Since $\Delta u_h$ is several times larger than $\Delta u_e$, because the gradient into the hot trapped electron population is flat, the absorption is closer to $\omega_{ce}$ than the emission. However, in addition the number density of trapped particles is at least one order of magnitude below that of the fast beam. Thus we may assume ${\tilde\alpha}\sim 10^{-4}$, and the maximum absorption rate can be estimated to be 
\begin{equation}
|\mathrm{Im}(\nu_a)|<5\times10^{-7}
\end{equation}
which is substantially less than the emission rate. Nevertheless, absorption of electromagnetic radiation is a feature which belongs to the physics of radiation from electron holes. If absorption becomes strong, its bandwidth may overlap the bandwidth of the emission band from electron holes and modify the emitted radiation; it may also cause narrow absorption bands on the background emission generated by the horseshoe-ring distribution passing the hole. 

\subsection*{\small{Transformation into the stationary observer's frame}}
The transformation of the above results into the frame of the observer must be done relativistically because in the electron-cyclotron maser we are dealing with a weak relativistic effect. This transformation implies a shift of the origin of the hole into the origin of the observer frame along the parallel velocity axis by the amount $V_h$ of the velocity of the hole. 

One may recall that we have not imposed the condition $k_\|=0$ of a vanishing parallel wave number in the calculation. At the contrary, in the hole frame of the (approximately) circular hole the resonance curve is a circle in which the parallel wave number drops out and thus plays no role. Consequently the emission is strictly perpendicular. 

We can now take advantage of this fact by considering the general weakly relativistic resonance condition with $k_\|\neq0$ in the observer's frame. It describes a resonant circle shifted along the parallel velocity direction by the (normalised to $c$) amount $k_\|c/\omega_{ce}$ \citep[cf., e.g.,][Chpt. 5, pp.120-125]{treumann1997}. This shift corresponds to the velocity of the hole thus providing a determination of the parallel wave number through the hole velocity
\begin{equation}\label{eq-kpara}
k_\|c/\omega_{ce}=V_h/c
\end{equation}
The above parallel wave number does not vanish for $V_h\neq0$. Hence, in the observer's frame the radiation emitted by the electron hole is oblique. This fact is of course nothing else but the well-known relativistic aberration effect by which the angle of propagation $\theta$ in the observer's frame is expressed through the angle of propagation $\theta'$ in the hole frame as
\begin{equation}
\tan\ \theta= \frac{\sin\ \theta'}{\gamma_h(\cos\ \theta'+\beta_h)}, \qquad \gamma_h=(1-\beta_h^2)^{-\frac{1}{2}}
\end{equation}
where $\beta_h\equiv V_h/c$. In the hole frame we have $\theta'=\frac{\pi}{2}$. Hence, the direction of radiation in the stationary frame becomes
\begin{equation}
\tan\ \theta=\beta_h^{-1}\sqrt{1-\beta_h^2}
\end{equation}
The larger $\beta_h$ the more will the direction of radiation deviate from the strictly perpendicular direction with respect to the ambient magnetic field. This is nothing else but the (weakly) relativistic beaming effect on the radiation by the moving radiation source which, in our case, is the electron hole. 

The change in frequency due to the displacement of the hole along the magnetic field is described by the relativistic Doppler effect which yields for the observed frequency of the radiation
\begin{equation}
\nu_{ce}=\nu_{ce}' \frac{\sqrt{1-\beta_h^2}}{1\pm\beta_h}, \qquad \nu_{ce}'\approx \nu_{ce}^m
\end{equation}
as function of the emitted frequency $\nu_{ce}'$. The sign $\pm$ in the denominator applies to separating and approaching holes, respectively. This is well known but shows that the emission frequency can shift to above the electron cyclotron frequency if the hole approaches the stationary observer. 

In the opposite case, for receding holes, the observed frequency decreases and drops farther below the electron cyclotron frequency.  Since $\nu_{ce}'\lesssim 1$, the observed frequency will clearly exceed the electron cyclotron frequency only if the hole approaches the origin of the observer frame at velocity 
\begin{equation}
\beta_h>{\scriptstyle\frac{1}{2}}(1-\nu_{ce}'^{2}) 
\end{equation}
which becomes possible for even modestly relativistic hole speeds because of the proximity to unity of the emitted frequency in the hole frame. Thus radiation from electron holes with frequency above the local electron cyclotron frequency should not be a surprise. It just requires propagation of the hole toward the observer at modestly relativistic speed.

\section{Discussion}

The present paper presents the head-on attempt of investigating the electromagnetic radiation that can be emitted by a single electron hole when it is generated in the presence of a ring-horseshoe distribution of the kind that is known to exist in the auroral-kilometric radiation-source region.
So far the auroral kilometric radiation has been considered to be the result of the global electron ring-horseshoe distribution itself which has been observed in the auroral auroral kilometric radiation-source region. In that case the global growth rates were of the order of $10^{-4}<\mathrm{Im}(\nu_{ce}^\mathit{ring})\sim \mathrm{few}\,10^{-3}$, {substantially larger than those estimated for single holes}.  

The persistently repeated observation of a distinct narrowband and fast-moving fine structure superposed on the auroral kilometric radiation spectrum, in particular when the temporal and spectral resolution increased, has always led to the suspicion that the comparably simple picture of generation of the auroral kilometric radiation solely by the global horseshoe distribution might be incomplete. The detection of electron holes in the same region combined with numerical simulations of the generation of electron holes provides ample reason for asking whether electron holes themselves could or could not contribute to the production of the observed spectral fine structure. In the simplified theory put forward in the present paper we have shown that this might be indeed the case. 

{We have provided arguments for the extended structure of the hole in velocity space. We argued that they necessarily have finite extension in $v_\perp$ if generated by the parallel Buneman instability. They will always become attracted by the horseshoe distribution because they alike positive charges which exchange momentum with the ring-electrons and become attracted by the negatively charged ring distribution. Holes in velocity space under conditions of strong magnetic fields, i.e. under conditions when $\omega_e\ll\omega_{ce}$, have been argued to be close to being circular (or spherical) in velocity space ($v_\|, v_\perp$), in overall agreement with observations in the auroral kilometric radiation source region. Entering into its centre, they deform the horseshoe sufficiently strongly such that steep perpendicular phase-space gradients are produced.}

{Such holes, as we have shown by direct calculation for a simple model, will necessarily generate band limited electron maser radiation with bandwidth comparable to the observed bandwidth of few kHz (or few \% of $\omega_{ce}$). In the hole frame this radiation is emitted in the X-mode in the direction perpendicular to the ambient magnetic field. The emitted wavelength is in the km-range and is comparable to the size of the hole in configuration space. Holes, hence, act like point sources of radiation while themselves moving at a susceptible fraction of the auroral electron beam along the magnetic field. Transforming the radiation into the spacecraft frame yields oblique radiation at Doppler shifted frequency. }  A number of interesting features is summarised as follows:

\subsection*{\small{Achievements}}

--- Since electron holes are localised entities not only in configuration but also in momentum space, maser theory can most conveniently be applied to them in the proper frame of the holes. For symmetry  reasons the radiation emitted in this case in the hole frame is about strictly into perpendicular direction which simplifies the calculation substantially. The radiation is therefore in the R-X mode.

\noindent
--- The radiation consists of two bands, an emission band resulting from the positive momentum space gradient at the outer boundary of the hole, and an absorption band resulting from the negative momentum space gradient of the interior trapped electron component in the hole. 

\noindent
--- The emitted bandwidth is, under auroral kilometric radiation conditions, of the order of $\sim1$ kHz  which is comparable to observed fine structure bandwidths. Absorption bandwidths are of similar order of magnitude. {The radiation frequency at $\sim450$ kHz corresponds to wavelengths of the order of $\sim700$ m which is the same order of magnitude as the spatial extension of the hole. Hence, the hole acts like a radiating source. This is, however, marginal because at such scales inhomogeneities of the plasma will come into play already. If this starts to be the case, refraction effects cannot be neglected anymore and the radiation might assume obliqueness and couple to other free space modes like the O-mode. Such problems have not been considered here. }

\noindent
--- Emitted frequencies in the proper frame of the hole are beneath the local non-relativistic electron-cyclotron frequency with the absorption frequencies being at higher frequency than the emitted frequencies.

\noindent
{--- Growth rates are less than the global ring-horseshoe growth rates $\sim 10^{-4}\omega_{ce}$ \citep{wulee1979,winglee1983,pritchett1986,louarn2006} by one order or more. Under the conservative conditions assumed here, they are of the order of $<10^{-5}\omega_{ce}$, while absorption rates are roughly two orders of magnitude less than this.} The growth rates produced by {single  electron holes are thus weak for the reason that few electrons only are involved into generation of radiation. On the other hand, as long as $\omega_{e}<\omega_{ce}$ the mechanism will work, Since the growth rate is proportional to the number density fraction $\alpha$ of electrons in the cold hole-accelerated beam and to the ratio $(\omega_e/\omega_{ce})^2$ it steeply grows with increasing density. Moreover, based on the high resolution electron data we argued that the growth rate of the ring distribution may be overestimated. In any case, maser radiation from holes will necessarily  generate fine structure in the auroral kilometric radiation spectrum. The magnitude of growth rates also} suggests that radiation will dominate absorption even if the emission and absorption bands overlap, which for moderately large beam velocities $U$ will not necessarily be the case.

\noindent
--- It is simple matter to transform the results to the fixed observer frame. Since the radiation source is not distributed over space and momentum space, the transformation from the hole to the observer frame along the magnetic field takes into account the parallel hole motion $V_h$. Relativistic transformation than turns the radiation, which was strictly perpendicular in the hole frame,  into oblique direction in the observer frame. At the same time the emitted frequency becomes Doppler shifted either down or up by the receding or approaching holes, respectively. Thus emission frequencies above the local electron cyclotron frequency become also possible for modestly relativistic parallel hole velocities. This is the case, because $V_h$ is of the order of the weakly relativistic ring-horseshoe beam velocity. 

\noindent
{--- The hole-radiation maser also offers a different interpretation of the observed drifts of the radiation fine structures as simple Doppler drifts. Decreasing frequencies in the spacecraft frame then suggest that the respective holes recede from the spacecraft while increasing frequencies imply that the holes approach the spacecraft. If we take the measured frequency drift (Fig. \ref{fig-rad}) of the fine-structures of $df/dt\sim -100$ kHz/s, measured parallel ring velocities $V_R\sim 2\times 10^4$ km/s, and bandwidth $\Delta f\sim 2$ kHz, the holes will have receded a distance of $\Delta z\sim 60-90$ km from the spacecraft in this observation along the magnetic field.}

{In all these cases, according to hole theory, the hole should move into the same direction like the downward ring electrons, i.e. downward. We may check this assumption with reference to Figure \ref{fig-akr-spectrum}, where no electron data are available. The bending of the fine structures in the upper radiation band observed at $\sim$6.5 s implies that the hole has passed close to the spacecraft after approaching it and afterwards recedes from it. At the bending which corresponds to closest approach the Doppler shift is zero. Thus the measured frequency is the proper emission frequency of the hole at $\sim442$ kHz which corresponds to an altitude of $h\sim1950$ km. The proper frequency change over the distance in frequency of $\sim 1$ kHz is not remarkable. Hence the drift in this case can indeed be interpreted as being due to Doppler shift in emission frequency of the downward moving radiating electron holes.}

All these results support the view that electron holes are a vital player in the electron-cyclotron maser theory of auroral kilometric radiation -- and possibly also in other objects, solar radio bursts, radiation from planets \citep{zarka2005,louarn2007,hess2009,mottez2010} like Jupiter and Saturn and astrophysical objects as possibly shocks \citep[e.g.][]{treumann2009,treumann2011}.  

\subsection*{\small{Weaknesses -Deficiencies}}

Finally we note a number of caveats. The first of these is that we have idealised the phase-space structure of an electron hole as being circular. 

The exact form of a hole perpendicular to the magnetic field is not known yet. Thus our assumption might be wrong by a factor varying between 1 and 10 in the ratio of the two main axes of the hole which implies that the circular structure would have to be replaced by an elliptical form either flat or elongated in the direction of the magnetic field (the latter being more probable in the strong auroral magnetic field case). Clearly this will affect the emitted radiation. It will, however, barely  substantially change the steepness of the gradient into perpendicular velocity direction. Thus growth rates will remain to be {similar}. What will change,  is the bandwidth of the radiation and absorption as well as the angle under which radiation is emitted since these depend on the phase-space geometry of the hole. For non-circular holes in phase space radiation will have an {intrinsic} oblique component, and the bandwidth will increase. These effects are important but could not be considered in this paper.

The second caveat concerns the consideration of only one hole. We estimated that the auroral kilometric source region contains a multitude of holes which all contribute to radiation and  absorption {at slightly different frequencies. These differences inhibit that emitted radiation from one hole will become amplified by passing another hole because the frequency of its emitted radiation is out of resonance with the other hole. It might accidentally happen to be so, however the more probable case is the former nonresonant case. It might also not be absorbed, however. Thus its power decays as the square of distance from the hole. Otherwise the radiation of the many holes contributes simply to the spectral width. This, however, is for perpendicular emission restricted to a narrow band near $\omega_{ce}$ only which unresolved then covers a band around $\omega_{ce}$ similar to those observed.} Since in real configuration space these holes are at different distance from the observation facility their individual radiation intensities decay with distance {from spacecraft}. A proper model should therefore include the spatial and velocity space distribution of holes in order to account for their global contribution.

Thirdly, electron holes move in space. Thus the emitted spectrum of a global distribution should change in time, an effect not considered here.

Fourth, electron holes themselves obey a violent internal {evolu}tion with time when evolving from Buneman instability or by any other means. They grow, merge, decay, saturate and so on. All these effects have not been considered here but in nature will necessarily  also affect the emission of radiation, causing time variations, drop-outs and other temporal effects. Since energy loss due to radiation should, however, not be extraordinarily large, it might not be expected that a model of radiation saturation by exhausting the energy of holes will be appropriate as all the other much more violent effects will much more strongly contribute to hole dynamics, saturation, stabilisation or even destruction of the holes. Being phase-space structures, holes cannot be indefinitely stable but will decay by internal processes of which radiation emission will presumably be one of the weakest while its observation will still provide information about the dynamics of the holes. {In fact, comparing the energy flux in the electrons in Figure \ref{fig-electrons-eh} with that in the auroral kilometric radiation in Figure \ref{fig-rad} we find that the latter is at least six orders of magnitude less than the former.}

{Finally we come to the most serious weakness of our theory, viz. the very small estimated growth rate of the hole-emitted radiation. This small growth rate implies that the e-folding time of radiated power flux ${\cal P}$ is $t\sim 5\times 10^4\omega_{ce}^{-1}\approx 2.4$ at the central $\sim480$ kHz in Figure \ref{fig-rad}. This time is thus of the order of $\sim 2$ s, much too long for generating the observed power flux of ${\cal P}\sim 10^{-11}$ W/m$^2$ within a few ms fraction of time. Assuming that the measured power would be emitted in 30 ms, comparable to the time resolution, then the thermal background of the radiation which would be amplified by the hole-cyclotron maser should have power flux ${\cal P}_\mathit{th}\sim 8.7\times10^{-12}$ W/m$^2$. This is impossible because the background flux is roughly five orders of magnitude lower (as read from the blue background noise in Figure \ref{fig-rad}) being around ${\cal P}_\mathit{th}^\mathit{obs}\sim <10^{-17}$ W/m$^2$.}

{Referring to the large number of electron holes involved, an escape could be searched in the number of holes contributing to one of the fine structures. This would require $\sim10^6$ holes in each fine structure behaving alike. In addition to the reservations noted earlier, this number is unreasonably high compared to the $\sim200$ holes passing across the spacecraft within one second along the magnetic field. If holes had transverse sizes of, say, 20 m, it would require that the holes occupying a fluxtube of $\sim 10$ km radius near spacecraft would all contribute to one fine structure in Fig. \ref{fig-rad}, which we believe is unrealistic.} 

{The missing ingredient  is a high resolution precision measurement of the electron energy flux which would resolve the holes in the horseshoe distribution. This requires measuring the electron distribution with time resolution of a few ms and energy resolution of, say, $\sim$10 eV at $>1$ keV. Clearly an instrumentation capable of this cannot be based on spacecraft spin for it must be able to measure the electron flux simultaneously in both the parallel and perpendicular directions to the magnetic field. }

{Here we are facing a real problem which we have not and could not overcome so far. Though everything in this theory seems to fall into its place, the quantitative estimate of the growth rate obtained is unsatisfactorily small such that it reproduces the observations only qualitatively and does not quantitatively account for the observed power in the auroral kilometric fine structures. We have provided an argument based on trapping of the radiation inside an electron hole for sufficiently long time, the life time of the hole against destruction by transverse instability. Trapping increases the amplification of the power by maintaining resonance between the radiation and the velocity space gradient. Estimated powers can in this way become comparable with observation. However, it remains unexplained how radiation could ultimately escape to free space.}

\conclusions
The {theoretical} results obtained in this {communication have been produced} by two methods, a qualitative consideration from the viewpoint of the observer frame and a quantitative approach in the hole frame. 

One might object that the two methods will not lead to the same result because the radiation in the observer's frame is genuinely oblique (the resonance circle becomes an ellipse shifted out of the origin) with $k_\|$ given in Eq. (\ref{eq-kpara}), while this obliqueness is not accounted for in the hole frame. This objection applies indeed only, when the hole is not circular as assumed here. {In the case of an elliptic electron hole the radiation emitted will be genuinely oblique and not constricted to be generated with its fundamental below the local electron cyclotron frequency.} But even in that case the calculation is easier to perform in the hole frame and subsequently transforming the result back into the observer's frame. This is possible because the hole represents kind of a point source which emits radiation of the order of the bandwidth provided by the size of its resonance circle. On the other hand, the qualitative results given for the circular hole in the observer's frame, i.e. the bandwidth of the emission and the resonance frequency, agree completely with those obtained in the hole frame for the reason that in the circular hole frame just one single resonance circle is needed in the steep gradient of the hole, while in the observer frame for strictly perpendicular emission a substantial number of resonance circles fill the interval between $R_\mathit{max}<R_\mathit{res}<R_\mathit{min}$, the maximum and minimum resonance radii, each of them crossing the steep perpendicular gradient in a very narrow interval only. {This is obvious from Figure \ref{fig-hole-rad-1}.} The line composed of all these narrow crossing segments of the resonance circles is identical to the resonance circle in the proper hole frame. Thus the estimate performed in the observer's frame agrees completely with the result obtained in the hole frame. Of course this should be so, because the relation between the two frames is just a relativistic translation.  

{The theory of electron hole radiation given here is simple and beautiful. It reproduces several features of the fine structure in the auroral kilometric radiation. However, the low growth rates are still in disagreement with the measured fluxes. This holds at least for the data of {\small FAST 1773} used in this paper where the plasma frequency in the upward current region was particularly low. In the case of {\small FAST 1750},shown in Figure \ref{fig-akr-spectrum}, the plasma density was somewhat larger yielding a one order of magnitude higher growth rate which, however, is still quite small and insufficient for reproducing the observed spectral radiation flux. One possibility of improving is amplification of radiation by wave trapping inside the electron hole. This has been discussed briefly. It brings up, however, some additional problems like modification of resonance and the general problem of radiation escape.}

{A possible way out of this dilemma could be attributing the generation of fine structure in the auroral kilometric radiation not to the upward but to the \emph{downward current} region. Here the densities and plasma frequencies are much larger while still  $\omega_e\ll\omega_{ce}$.  From observation it is known that phase space holes including electron holes exist in multitude here as well. However, the different physics of the downward current region requires a different treatment of holes and generation of radiation.}

\begin{acknowledgements}
This research was part of an occasional Visiting Scientist Programme in 2006/2007 at ISSI, Bern. RT thankfully recognises the assistance of the ISSI librarians, Andrea Fischer and Irmela Schweizer. He highly appreciates the encouragement of Andr\'e Balogh, Director emeritus at ISSI. The observational data used in this text have been obtained by the FAST spacecraft within the France-University of California at Berkeley cooperation. Figure \ref{fig-over} (and some other data as well) have previously been published in the cited papers. They are reprinted here with changes and with the permission of the American Geophysical Union. RT thanks the two referees, which unfortunately remained anonymous, for forcing him to include and to properly analyse a substantial amount of data. We also thank the referees for their critical as well as their well justified sceptical remarks in particular on the value of the growth rate and the general importance of electron holes in the generation of radiation. Some of the data work was done ten years ago with the help of Edita Georgescu whose efforts are thankfully acknowledged at this place.\end{acknowledgements}

\end{document}